\documentclass[aps,prl,reprint,superscriptaddress,letterpaper]{revtex4-1}
\usepackage[english]{babel}
\usepackage{ucs}
\usepackage[utf8x]{inputenc}
\usepackage{amsmath}
\usepackage{amsfonts}
\usepackage{amssymb}
\usepackage{graphicx}
\usepackage{bm}
\usepackage{bbm}
\usepackage{empheq}
\usepackage{hyperref}

\begin{document}
\title{Observation of classical-quantum crossover of $1/f$ flux noise\\and its paramagnetic temperature dependence}

\author{C. M. Quintana}
\affiliation{Department of Physics, University of California, Santa Barbara, California 93106, USA}
\author{Yu Chen}
\affiliation{Google Inc., Santa Barbara, California 93117, USA}
\author{D. Sank}
\affiliation{Google Inc., Santa Barbara, California 93117, USA}
\author{A. G. Petukhov}
\affiliation{NASA Ames Research Center, Moffett Field, California 94035, USA}
\author{T. C. White}
\affiliation{Google Inc., Santa Barbara, California 93117, USA}
\author{Dvir Kafri}
\affiliation{Google Inc., Venice, CA 90291, USA}
\author{B. Chiaro}
\affiliation{Department of Physics, University of California, Santa Barbara, California 93106, USA}
\author{A. Megrant}
\affiliation{Google Inc., Santa Barbara, California 93117, USA}

\author{R. Barends}
\affiliation{Google Inc., Santa Barbara, California 93117, USA}
\author{B. Campbell}
\affiliation{Department of Physics, University of California, Santa Barbara, California 93106, USA}
\author{Z. Chen}
\affiliation{Department of Physics, University of California, Santa Barbara, California 93106, USA}
\author{A. Dunsworth}
\affiliation{Department of Physics, University of California, Santa Barbara, California 93106, USA}
\author{A. G. Fowler}
\affiliation{Google Inc., Santa Barbara, California 93117, USA}
\author{R. Graff}
\affiliation{Google Inc., Santa Barbara, California 93117, USA}
\author{E. Jeffrey}
\affiliation{Google Inc., Santa Barbara, California 93117, USA}
\author{J. Kelly}
\affiliation{Google Inc., Santa Barbara, California 93117, USA}
\author{E. Lucero}
\affiliation{Google Inc., Santa Barbara, California 93117, USA}
\author{J. Y. Mutus}
\affiliation{Google Inc., Santa Barbara, California 93117, USA}
\author{M. Neeley}
\affiliation{Google Inc., Santa Barbara, California 93117, USA}
\author{C. Neill}
\affiliation{Department of Physics, University of California, Santa Barbara, California 93106, USA}
\author{P. J. J. O'Malley}
\affiliation{Department of Physics, University of California, Santa Barbara, California 93106, USA}
\author{P. Roushan}
\affiliation{Google Inc., Santa Barbara, California 93117, USA}
\author{A. Shabani}
\affiliation{Google Inc., Venice, CA 90291, USA}
\author{V. N. Smelyanskiy}
\affiliation{Google Inc., Venice, CA 90291, USA}
\author{A. Vainsencher}
\affiliation{Google Inc., Santa Barbara, California 93117, USA}
\author{J. Wenner}
\affiliation{Department of Physics, University of California, Santa Barbara, California 93106, USA}
\author{H. Neven}
\affiliation{Google Inc., Venice, CA 90291, USA}
\author{John M. Martinis}
\affiliation{Department of Physics, University of California, Santa Barbara, California 93106, USA}
\affiliation{Google Inc., Santa Barbara, California 93117, USA}
\date{\today}

\begin{abstract}
By analyzing the dissipative dynamics of a tunable gap flux qubit, we extract both sides of its two-sided environmental flux noise spectral density over a range of frequencies around $2k_BT/h \approx 1\,\rm{GHz}$, allowing for the observation of a classical-quantum crossover. Below the crossover point, the symmetric noise component follows a $1/f$ power law that matches the magnitude of the $1/f$ noise near $1\,{\rm{Hz}}$. The antisymmetric component displays a $1/T$ dependence below $100\,\rm{mK}$, providing dynamical evidence for a paramagnetic environment. Extrapolating the two-sided spectrum predicts the linewidth and reorganization energy of incoherent resonant tunneling between flux qubit wells.
\end{abstract}


\maketitle
The ubiquity of low-frequency magnetic flux noise in superconducting circuits has been well-known for decades \cite{Clarke1976, Koch1983, Wellstood1987, Yoshihara2006, Bialczak2007, Sendelbach2008, Lanting2009, Anton2013}, making it perhaps surprising that its microscopic origin has yet to be determined. Understanding the origin of flux noise will be crucial to reduce it \cite{Kumar2016}, a key task for the advancement of technologies based on flux-sensitive superconducting circuits. In general, flux noise causes parameter drift and loss of phase coherence.
In quantum annealing, low-frequency flux noise limits the number of qubits that can coherently tunnel, and a dissipative flux environment can suppress incoherent quantum tunneling \cite{Leggett1984, Esteve1986, Amin2008, Harris2008, Boixo2016}.

In the past decade, superconducting qubits have extended the measurement of flux noise to increasingly wider frequency ranges, showing a $1/f^{\alpha(f)}$ power spectrum from $f\approx10^{-5}\,\rm{Hz}$ to $f\approx 1\,\rm{GHz}$ with $\alpha$ close to 1 at low temperatures \cite{Bialczak2007, Bylander2011, Sank2012, Yan2012, Slichter2012, Yan2013, Yan2016}. Although previous frequency-resolved measurements of this noise have used a variety of experimental methods, they have extracted only a single quantity, the symmetrized spectrum $S^+_\Phi(f)$, to describe the noise versus frequency. However, as is well known, a quantum environment can generate different noise spectra at positive and negative frequencies, $S_\Phi(-f) \ne S_\Phi(f)$, with $S^+_\Phi(f) \equiv S_\Phi(f) + S_\Phi(-f)$. Physically, the asymmetric quantum part $S^-_\Phi(f) \equiv S_\Phi(f) - S_\Phi(-f)$ is a measure of the density of states of the environment. If the environment is in thermal equilibrium, the two spectra are related through the fluctuation-dissipation theorem, $S_\Phi^+(f) = \coth(h f/2k_B T) S_\Phi^-(f)$ \cite{Devoret1997}. In the classical regime $f \ll 2 k_B T/h$, $S_\Phi^+$ dominates and $S_\Phi^-$ is negligible unless one is sensitive to small levels of dissipation. Conversely, in the quantum limit of high frequencies one has $S^+_\Phi \approx S^-_\Phi$, meaning a single spectrum describes the environment unless one is sensitive to small levels of thermal noise. But when $f$ is of order $k_B T/h$ or if there is non-equilibrium noise, a single spectrum no longer characterizes the environment. There is recent evidence that $1/f$ noise contributes to qubit relaxation in the quantum regime \cite{Bylander2011, Yan2016}, but there has been no frequency-resolved experimental distinction between $S_\Phi^+$ and $S_\Phi^-$ or between equilibrium and non-equilibrium noise. Furthermore, the transition from classical to quantum flux noise has yet to be observed or understood.

Experimentally, obtaining $S_\Phi^-(f)$ in the classical or crossover regime is challenging, as it requires high fidelity readout of the qubit population over a range of frequencies lower than the temperature, outside typical superconducting qubit operating conditions. In this Letter, we implement a tunable gap flux qubit called the ``fluxmon'' with a measurement scheme that allows us to study this physics. We extract the full two-sided flux noise spectrum over a range of frequencies containing $2k_B T/h\sim1\,\rm{GHz}$. We observe a classical-quantum crossover, which we find to coincide with a transition from $1/f$ to quasi-ohmic dissipation. Remarkably, we find that $S_\Phi^+(f)\propto 1/f$ at $1\,\rm{GHz}$, with a magnitude closely matching that extrapolated from the $1/f$ noise below $1\,\rm{Hz}$, 10 orders of magnitude away. The strength of the $1/f$ noise at high and low frequencies changes similarly between samples \cite{supplement}, providing evidence that they may originate from the same physical source. Below the crossover point we find that the environment is close to thermal equilibrium. We measure the temperature dependence of $S^\pm_\Phi(f)$, and discover a paramagnetic $1/T$ scaling in $S_\Phi^-$. Finally, we show that the small but non-zero $S_\Phi^-$ in the classical regime has an important effect by using it to predict the classical reorganization energy associated with incoherent tunneling between flux qubit wells \cite{Harris2008}, a crucial quantity for modeling the performance of a quantum annealer.

\begin{figure}[t]
\begin{centering}
\includegraphics[width=1\linewidth]{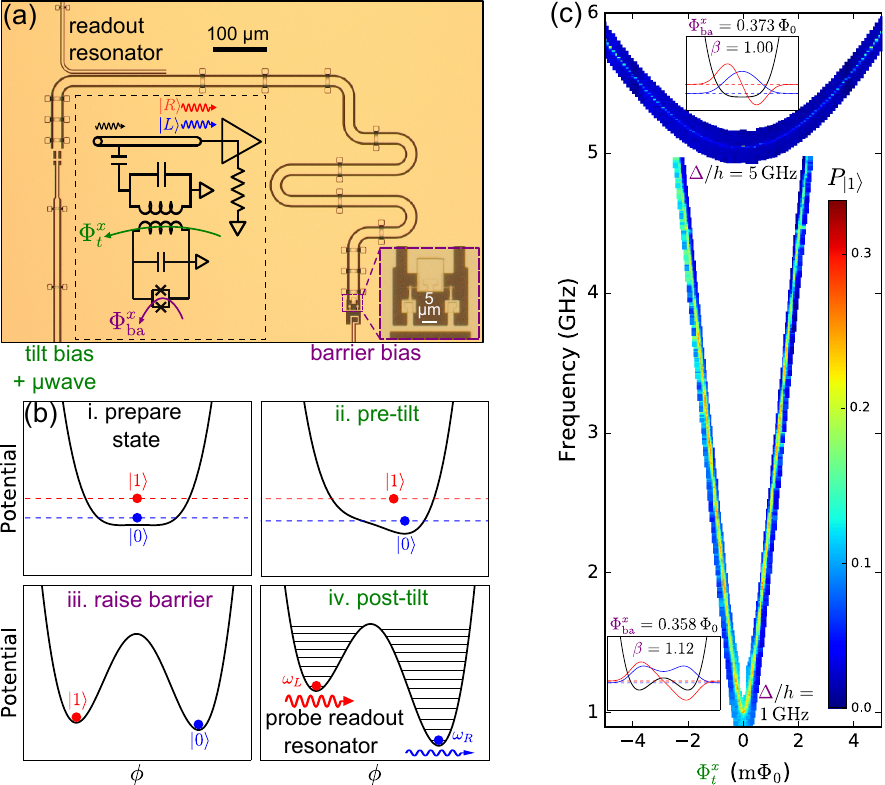}
\par\end{centering}
\caption{\footnotesize (color online) (a) Optical micrograph of Al/sapphire fluxmon qubit and bias lines, inductively coupled to a CPW readout resonator. (b) Four stages of measurement, involving double-well projection followed by dispersive readout. (c) Qubit spectroscopy versus frequency and $\Phi_t^x$ for two values of $\Delta(\Phi_{\rm{ba}}^x)$.}
\label{figure:fig1}
\end{figure}

The coplanar waveguide (CPW)-based fluxmon qubit was designed with a long loop and appreciable persistent current to achieve strong inductive coupling to many qubits at once for quantum annealing. This also makes it a sensitive tool for studying flux noise. As illustrated in Fig. \ref{figure:fig1}(a), the fluxmon consists of a CPW segment shorted on one end and shunted with a DC SQUID on the other. The CPW has distributed inductance and capacitance per unit length $\mathcal{L}$ and $\mathcal{C}$, with length $\ell$ such that at frequencies below its $\lambda/4$ resonance,
it acts as a lumped-element parallel inductance $L=\mathcal{L}\ell \approx 600\,\rm{pH}$ and capacitance $C=\mathcal{C}\ell/3\approx 100 \,\rm{fF}$. Shunting it with a DC SQUID adds a tunable nonlinear term to the potential energy, allowing the the potential to be varied from a harmonic single well to a double-well regime characteristic of a flux qubit \cite{Mooij1999}. The fluxmon's main ``loop'' is the CPW mode of current flow. The potential's ``tilt'' bias is tuned by an external flux $\Phi_t^x$ through the CPW, while the barrier height is tuned via the DC SQUID external flux $\Phi^x_{\rm{ba}}$, yielding a tunable Josephson term parameterized through the variable $\beta = (2\pi/\Phi_0)^2 E_J L = \beta_{\rm{max}}\cos(\pi\Phi_{\rm{ba}}^x/\Phi_0)$, where $\beta_{\rm{\max}} \approx 2.5$. Grounds are connected throughout the circuit with Al airbridges \cite{Chen2014} to control linear crosstalk and coupling to spurious modes.

We implement a variation of dispersive readout inspired by phase qubits \cite{Steffen2010Readout, Chen2012}. After performing qubit state manipulations with microwave pulses in the single-well regime, we adiabatically project the qubit energy states into the stable left and right wells of a double-well potential with a raised barrier, as illustrated in Fig. \ref{figure:fig1}(b). A large tilt is then applied to induce different well frequencies that are detected through a dispersively coupled resonator, yielding a single-shot separation fidelity of $>99.9 \%$ and a $|1\rangle$ state fidelity of $\sim 97\%$ limited by energy decay during the projection.

\begin{figure*}[t!]
\centering
\includegraphics[width=2.0\columnwidth]{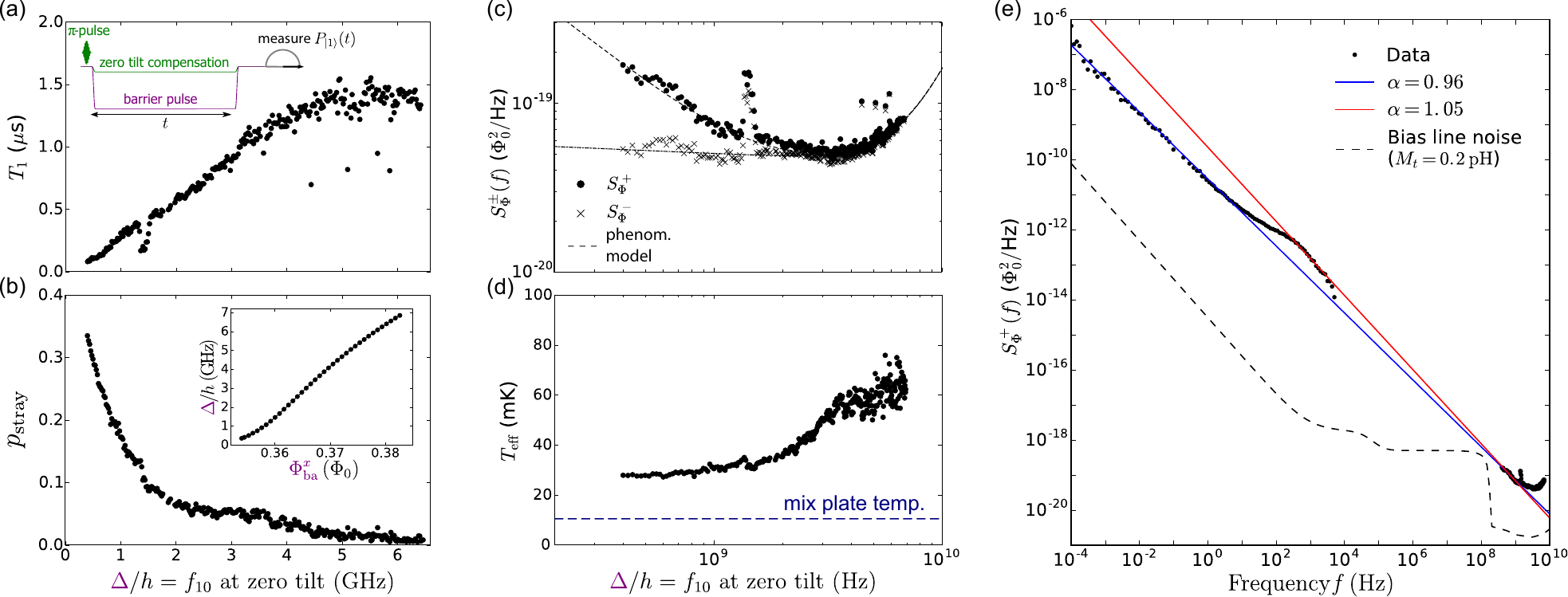}
\caption{\footnotesize (a) $T_1$ versus $f_{10}$ at zero tilt measured with swap spectroscopy. Inset illustrates the pulse sequence. (b) $p_{\rm{stray}}$ versus $f_{10}$ at zero tilt. Inset: $f_{10}$ at zero tilt versus barrier bias. (c) $S_\Phi^\pm(f)$ extracted from (a) and (b) and numerical evaluation of $\langle 0|\hat{\Phi}|1\rangle$. (d) Effective temperature $T_{\rm{eff}}$ of the data in (b). (e) Low-frequency quasistatic flux noise \cite{supplement} together with $S_\Phi^+(f)$ from (c) on the same axes.}
\label{figure:fig2}
\end{figure*}
At zero tilt bias, the potential energy function is symmetric, meaning the energy eigenstates $|0\rangle$ and $|1\rangle$ have even and odd parity. Using this parity basis in the two-level approximation (not too far from zero tilt bias), the flux qubit Hamiltonian can be written as $\hat{H} = -\frac{1}{2}[\Delta(\Phi_{\rm{ba}}^x) \sigma_z + \varepsilon(\Phi_t^x) \sigma_x]$, where the gap $\Delta/h$ is the qubit frequency at zero tilt, and $\varepsilon = 2I_p\Phi_t^x$. Here, $I_p \equiv \frac{1}{L}|\langle 0 | \hat{\Phi} | 1\rangle|\sim 0.5\,\mu\rm{A}$ is the ``persistent current.'' At zero tilt, flux noise in the main qubit loop (i.e., in $\varepsilon$) at the qubit frequency induces incoherent transitions between energy eigenstates according to Fermi's golden rule \cite{Schoelkopf2003}: $\Gamma_{\downarrow/\uparrow} = \frac{1}{\hbar^2}\frac{1}{L^2}|\langle 0|\hat{\Phi}|1\rangle|^2 S_\Phi(\pm f_{10})$. This implies a two-rate equation in which the qubit relaxes to a steady state population $p_{\rm{stray}} = \Gamma_\uparrow/(\Gamma_\downarrow + \Gamma_\uparrow)$ at a rate $1/T_1 = \Gamma_\downarrow + \Gamma_\uparrow$. From $T_1$ and $p_{\rm{stray}}$ we can extract both $S_\Phi^+$ and $S_\Phi^-$:
\begin{align}
S^+_\Phi(f_{10}) &= \frac{\hbar^2 L^2}{T_1|\langle 0 | \hat{\Phi} | 1 \rangle |^2}\label{main_equation},\\ S_\Phi^-(f_{10}) &= [1 - 2 p_{\rm{stray}}(f_{10})] S_\Phi^+(f_{10}).\label{main_equation_2}
\end{align}

We measure $T_1$ and $p_{\rm{stray}}$ vs. $f_{10}$ at zero tilt bias by varying the barrier bias, and use the data to extract $S_\Phi^\pm(f)$ after numerically computing $\langle 0 |\hat{\Phi}|1\rangle$. The ability to tune $\Delta$ as in Fig. \ref{figure:fig1}(c) is crucial, as it allows us to vary frequency while remaining at zero tilt so as to be sensitive only to transverse noise in $\varepsilon$. It also allows measurement over a wide frequency range within a valid two-level approximation. We use the method of ``swap spectroscopy'' \cite{Mariantoni2011} [Fig. \ref{figure:fig2}(a) inset]: first, with $f_{10} \approx 5\,\rm{GHz}$ ($\beta \approx 1$), the qubit is excited using a $\pi$-pulse, and then a barrier pulse detunes $f_{10}$ to a different frequency. The detuning pulse is slow enough to be adiabatic, yet much shorter than $T_1$. During the barrier pulse, a compensating tilt bias pulse is applied to correct for crosstalk, keeping the qubit at zero tilt. We wait for a variable time $t$ at the detuned bias point, before tuning back to the original bias and performing readout. After measuring $p_{|1\rangle}(t)$, we extract $T_1$ and $p_{\rm{stray}}$ by fitting to $p_{|1\rangle}(t) = p_0\exp(-t/T_1) + p_{\rm{stray}}$. A typical dataset for $T_1$ and $p_{\rm{stray}}$ vs. $\Delta/h$ is shown in Fig. \ref{figure:fig2} (a) and (b).

A typical dataset converted to spectral densities [Eqs. (\ref{main_equation}) and (\ref{main_equation_2})] is shown in Fig. \ref{figure:fig2}(c). In addition, the extracted effective temperature $T_{\rm{eff}}$ is plotted in Fig. \ref{figure:fig2}(d), where $\exp(-h f_{10}/k_B T_{\rm{eff}}) \equiv \Gamma_\uparrow/\Gamma_\downarrow$. We observe several interesting features. First, below $\sim1\,\rm{GHz}$, $S_\Phi^+(f)$ follows a $1/f^\alpha$ law, with $\alpha\approx1$. Remarkably, in Fig. \ref{figure:fig2}(e) we find that extrapolating this power law to frequencies below $1\,\rm{Hz}$ predicts the magnitude of the quasistatic $1/f$ noise \cite{supplement} surprisingly closely. In thermal equilibrium, $S_\Phi^-(f)$ should scale as $f^{1 - \alpha}$ at low frequencies, meaning a constant $S_\Phi^-(f)$ for $1/f$ noise, which is roughly what we observe. At frequencies below the classical-quantum crossover, the noise appears to be described by a single $T_{\rm{eff}}\approx 30\,\rm{mK}$, suggesting thermal equilibrium of the low-frequency environment, but with $T > T_{\rm{fridge}}$ [Fig. \ref{figure:fig2}(d)]. We note there is a feature of unknown origin at $\sim 1.4\,\rm{GHz}$ \cite{supplement}.

Secondly, we find that the classical-quantum crossover is accompanied by a transition from $1/f^\alpha$ to super-ohmic dissipation, meaning noise for which $S^+_\Phi(f) \approx S^-_\Phi(f) \propto f^\gamma$ with $\gamma \ge 1$. 
If we fit to the phenomenological thermodynamic model $S_\Phi(\omega) = A\omega/|\omega|^{\alpha} [1 + \coth{(\hbar \omega/[2 k_B T_A])}] + B \omega |\omega|^{\gamma-1} [1 + \coth{(\hbar \omega/[2 k_B T_B])}]$, then $\alpha = 1.05$ and $\gamma$ between $2.5$ and $3$ fits our data best. Purcell decay is negligible. We note that $\gamma=3$ gives a frequency dependence for $T_1$ that is indistinguishable from ohmic charge noise ($S^-_Q\propto \omega$), the high-frequency model used in Ref. \cite{Yan2016}. Dielectric loss \cite{Martinis2005} would give $\gamma=2$, but based on similarly fabricated Xmon qubits and airbridges we estimate that that dielectric loss should limit $T_1$ to $\sim 20\,\mu$s at $5\,\rm{GHz}$. Allowing for a finite high-frequency cutoff for the $1/f^\alpha$ noise could yield a different best-fit $\gamma\approx1.5$ \cite{supplement}. If we instead simply ignore the $1/f$ part and only fit the data above $\sim 3.5\,\rm{GHz}$, an ohmic flux noise model with $\gamma\approx1$ describes the data reasonably well, with the net dissipation represented by a frequency-independent parallel resistance $R\approx 20\,{\rm{M}}\Omega$. Dissipation from the tilt flux bias line would similarly have $\gamma=1$, but with $T_1\approx40\,\mu\rm{s}$. Since the observed level of dissipation is not seen in the Xmon transmon qubit \cite{Barends2013}, which is comprised of a similar capacitance and Josephson inductance (albeit with a critical current density 10 times smaller) but negligible geometric inductance, we hypothesize that the quasi-ohmic noise is intrinsic to the magnetic degree of freedom. The presence of unexplained ohmic dissipation ($\gamma=1$) was seen in microstrip-based flux qubits \cite{Lanting2011}, but with a much stronger magnitude ($R\approx 20\,\rm{k}\Omega$). Given that the extracted $\gamma$ depends on whether or not the $1/f$ noise is included in the model, it could be that this earlier result was the combined effect of $1/f$ and super-ohmic dissipation. We also note that at high frequencies, $T_{\rm{eff}}$ ranges from $50 - 80\,\rm{mK}$, meaning a higher $T$ for the quasi-ohmic bath and/or the presence of non-equilibrium noise, making it difficult to model.

\begin{figure}[t!]
\begin{centering}
\includegraphics[width=.89\linewidth]{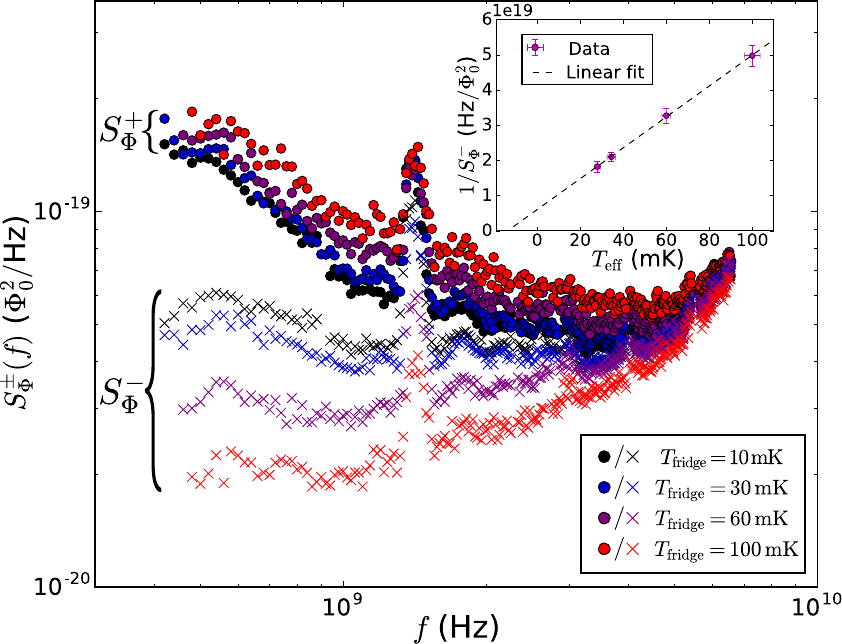}
\par\end{centering}
\caption{\footnotesize Temperature dependence of $S_\Phi^\pm$. $S_\Phi^-(f=500\,\rm{MHz})$ shows a $1/(T_{\rm{eff}} + T_0)$ dependence, as explicitly plotted in the inset.}
\label{figure:fig3}
\end{figure}

We investigate the nature of the $1/f$ noise further by looking at its temperature dependence. The temperature-independence of the classical low-frequency noise at millikelvin temperatures and the $1/T$ dependence of the static susceptibility in SQUIDs are evidence of a paramagnetic origin \cite{Sendelbach2008} for the $1/f$ flux noise. Here, we uncover dynamical evidence for this conclusion. As shown in Fig. \ref{figure:fig3}, we find that $S^-_\Phi$ displays an approximately $1/T$ dependence below the classical-quantum crossover point, which through the fluctuation-dissipation theorem implies that $\chi''(\omega)$, the imaginary (absorptive) part of the environment's dynamic susceptibility $\chi(\omega) = \chi'(\omega) + i\chi''(\omega)$, has a paramagnetic $1/T$ scaling (see \cite{supplement}). In comparison, $S^+_\Phi$ displays only a very slight temperature dependence, consistent with the fact that we see no measurable temperature dependence in the quasistatic flux noise ($f < 10\,\rm{kHz}$) \cite{supplement}. In the inset to Fig. \ref{figure:fig3} we explicitly plot $1/S^-(500\,\rm{MHz})$ versus $T_{\rm{eff}}$, implying that $\chi''(\omega,T) \propto 1/(T + T_0)$, with $T_0\approx 10\,\rm{mK}$. This functional form might be taken as evidence for paramagnetic spins that would behave antiferromagnetically at lower temperatures. We also note that a model with a temperature-dependent high-frequency cutoff of a few $k_B T_{\rm{eff}}/h$ (consistent with spin-phonon or spin-spin interactions) fits the crossover region vs. $f$ and $T$ somewhat better than one with a fixed or infinite cutoff (see \cite{supplement}).

$\\$

Finally, we show that the measured $S^-_\Phi$ extends deep into the classical regime by performing an experiment closely tied to quantum annealing. We look at the effect of dissipation on incoherent macroscopic resonant tunneling (MRT) between the lowest states of the left and right flux qubit wells. In the regime of large $\beta$, the tunnel coupling $\Delta/h$ is much smaller than the linewidth of the energy levels, meaning that quantum tunneling will be incoherent. For the level of damping and dephasing in our system, at temperatures below $T_{\rm{cr}} \approx 200\,\rm{mK} $ the escape rate from one well to the other near resonance should be dominated by quantum tunneling \cite{Buttiker1983, Devoret1985, Martinis1987, Amin2008}. Rewriting the result of \cite{Amin2008}, the tunneling rate as a function of tilt bias energy $\varepsilon$ is predicted to be
\begin{equation}
\Gamma_{L\to R}(\varepsilon) = \frac{\Delta^2}{4\hbar^2} \int_0^\infty dt\,e^{I_A(t)} \cos[\varepsilon t/\hbar - I_B(t)],\label{MRT}
\end{equation}
where $I_A(t) = \int_{f_l}^{f_h} df\,\frac{(2I_p)^2}{(h f)^2}S^+_\Phi(f) \cos(2\pi f t)$, $I_B(t) = \int_{f_l}^{f_h} df\,\frac{(2 I_p)^2}{(h f)^2} S^-_\Phi(f)\sin(2\pi f t)$, and $f_l$ and $f_h$ are appropriately chosen low and high frequency cutoffs. Assuming the integrated noise is dominated by frequencies smaller than the resonant tunneling linewidth $W/h$, then near its peak the tunneling rate  Eq. (\ref{MRT}) can be approximated as
$\Gamma(\varepsilon) = \sqrt{\frac{\pi}{8}} \frac{\Delta^2}{\hbar W} \exp\left[-\frac{(\varepsilon - \varepsilon_p)^2}{2 W^2}\right]$, where $W^2 = 4 I_p^2 \int_{f_l}^{f_h} df\,S^+_\Phi(f)$ and $\varepsilon_p = 4 I_p^2 \int_{f_l}^{f_h} df\,S^-_\Phi(f)/(h f)$ \cite{Amin2008}. It is therefore possible to observe the integrated effect of $S^-_\Phi$ in the deep classical regime by looking at the offset $\varepsilon_p$ for the maximum tunneling rate, which physically corresponds to the reorganization energy of the environment that must be absorbed upon tunneling.

\begin{figure}[h!]
\begin{centering}
\includegraphics[width=.82\linewidth]{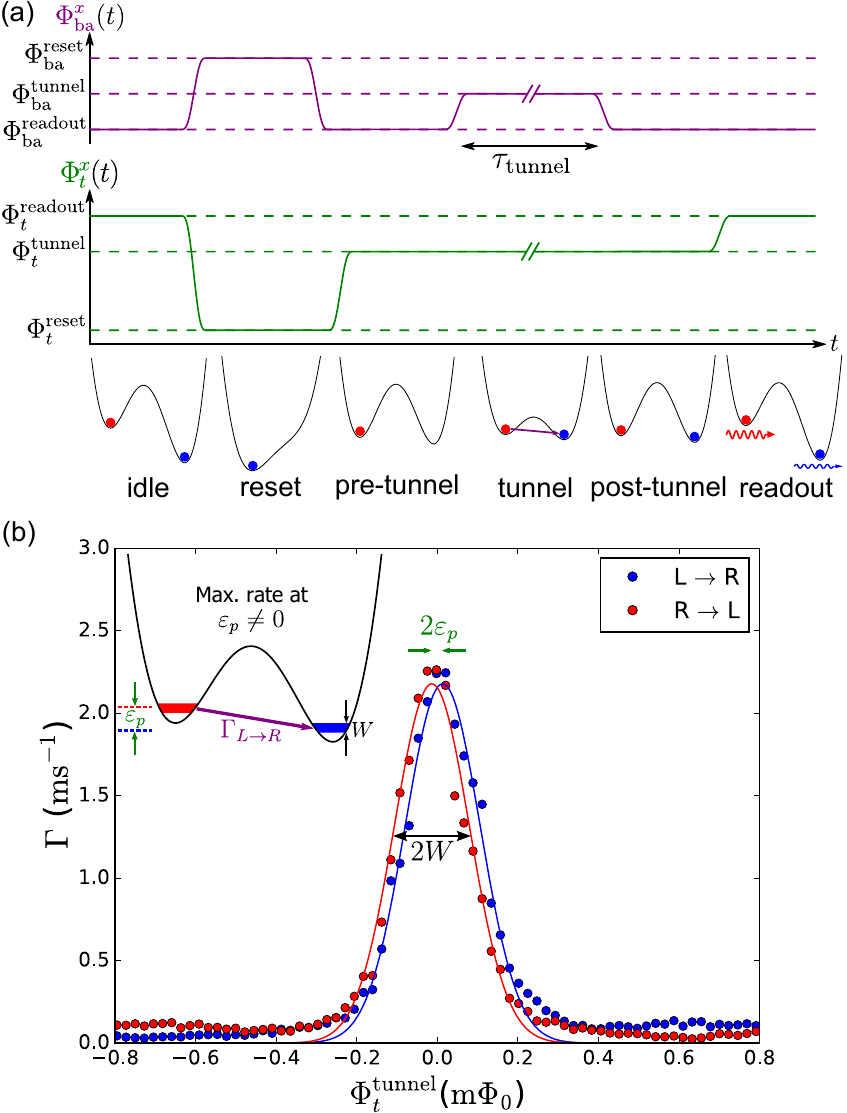}
\par\end{centering}
\caption{\footnotesize (a) Pulse sequence for the MRT experiment. (b) MRT data with $\Delta/h\approx 0.5\,\rm{MHz}$, showing a small offset in the tilt bias that maximizes tunneling, consistent with the measured $S_\Phi^-$.}
\label{figure:fig4}
\end{figure}

We measure $\Gamma(\varepsilon)$ using the pulse sequence in Fig. \ref{figure:fig4}(a). We prepare either the left or right well ground state with a high barrier, and as a function of tilt bias we lower the barrier to $\beta \approx 1.5$ ($\Delta/h \approx 1\,\rm{MHz}$) and measure the incoherent tunneling rate to the other well. A typical dataset is shown in Fig. \ref{figure:fig4}(b). Fitting the tunneling peaks to Gaussians, over multiple datasets we extract $\varepsilon_p/(2 I_p) = 7\pm 3\,\mu\Phi_0$ and $W/(2 I_p) = 80\pm20\,\mu\Phi_0$. Above base temperature $W$ is not changed within the margin of error, but $\varepsilon_p$ becomes too small to reliably measure.

We can compare $W$ and $\varepsilon_p$ to what we would expect from directly integrating $S^\pm_\Phi(f)$, interpolating a $1/f$ power law between the noise measured at low and high frequencies [Fig. \ref{figure:fig2}(e)]. Including the ohmic noise leaves the tunneling rate virtually unaffected near the peak even when integrating (\ref{MRT}) up to $f_h = 10\,\rm{GHz}$, the oscillation frequency of the inverted potential barrier, the natural high frequency cutoff \cite{supplement, Buttiker1982, Esteve1989}. The natural low-frequency cutoff for $W$ and $\varepsilon_p$ is the peak tunneling rate itself, $\sim10^{3}\,\rm{Hz}$. However, there is additional broadening of $W$ due to quasistatic noise averaged over experimental repetitions, which amounts to extending the low frequency cutoff for $W$ down to the inverse total data acquisition time \cite{supplement}. 
Using these cutoffs we predict $W/(2 I_p)\approx 50\,\mu\Phi_0$ and $\varepsilon_p/(2I_p)\approx 4\,\mu\Phi_0$, within a factor of two of the measured values. For low-frequency noise in thermal equilibrium, one would expect $T = W^2/(2 k_B \varepsilon_p)$ \cite{Harris2008}. Plugging in our measured $W$ and $\varepsilon_p$ yields $T_{\rm{eff}} \approx 60\,\rm{mK}$, significantly higher than the $30\,\rm{mK}$ deduced in Fig. \ref{figure:fig2}(d). However, this may be explained by the extra broadening of $W$ from quasistatic noise. Subtracting out our estimation of the contribution from this frequency range yields instead $20\,\rm{mK}$, qualitatively supporting this explanation.

In conclusion, we have used the fluxmon to measure flux noise over a range of frequencies about $2k_B T/h$, separately extracting the symmetric and antisymmetric components $S_\Phi^\pm(f)$ and observing the classical-quantum crossover. We find that $S_\Phi^-$ displays a paramagnetic temperature dependence below the crossover, and that $S_\Phi^+$ follows a $1/f$ power law whose magnitude is consistent with that of the $1/f$ flux noise near $1\,\rm{Hz}$. The fact that the noise spectrum has a $1/f$ shape near the crossover indicates that the underlying magnetic fluctuators have a distribution of relaxation times that extends to at least $1\,\rm{GHz}$, possibly hinting towards spin clustering as opposed to spin diffusion \cite{Faoro2008, Wang2015, supplement}, which would also be consistent with the correlated low-frequency inductance fluctuations observed in SQUIDs that were postulated to arise from fluctuations in spin cluster relaxation times \cite{Sendelbach2009}. Recent evidence \cite{Kumar2016} that adsorbed molecular O$_2$ (spin-1) may play a dominant role in flux noise could also support this conclusion, as spin-orbit induced magnetic anisotropy could break conservation of total spin and allow clusters to locally transfer energy and angular momentum to the lattice. Finally, we showed that the measured noise and dissipation can approximately predict incoherent quantum tunneling rates between flux qubit wells, which has direct implications for quantum annealing applications.

\begin{acknowledgments}
We thank Sergio Boixo, Robert McDermott, Fedir Vasko, Mostafa Khezri, and Fei Yan for insightful discussions. This work was supported by Google. C. Q. and Z. C. acknowledge support from the National Science Foundation Graduate Research Fellowship under Grant No. DGE-1144085. A. P. acknowledges support from the Air Force Research Laboratory (AFRL) Information Directorate under grant F4HBKC4162G001. Devices were made at the UC Santa Barbara Nanofabrication Facility, a part of the NSF-funded National Nanotechnology Infrastructure Network.
\end{acknowledgments}

%

\end{document}


\title{Supplemental Material for ``Observation of classical-quantum crossover of\\ $1/f$ flux noise and its paramagnetic temperature dependence''}

\author{C. M. Quintana}
\affiliation{Department of Physics, University of California, Santa Barbara, California 93106, USA}
\author{Yu Chen}
\affiliation{Google Inc., Santa Barbara, California 93117, USA}
\author{D. Sank}
\affiliation{Google Inc., Santa Barbara, California 93117, USA}
\author{A. G. Petukhov}
\affiliation{NASA Ames Research Center, Moffett Field, California 94035, USA}
\author{T. C. White}
\affiliation{Google Inc., Santa Barbara, California 93117, USA}
\author{Dvir Kafri}
\affiliation{Google Inc., Venice, CA 90291, USA}
\author{B. Chiaro}
\affiliation{Department of Physics, University of California, Santa Barbara, California 93106, USA}
\author{A. Megrant}
\affiliation{Google Inc., Santa Barbara, California 93117, USA}

\author{R. Barends}
\affiliation{Google Inc., Santa Barbara, California 93117, USA}
\author{B. Campbell}
\affiliation{Department of Physics, University of California, Santa Barbara, California 93106, USA}
\author{Z. Chen}
\affiliation{Department of Physics, University of California, Santa Barbara, California 93106, USA}
\author{A. Dunsworth}
\affiliation{Department of Physics, University of California, Santa Barbara, California 93106, USA}
\author{A. G. Fowler}
\affiliation{Google Inc., Santa Barbara, California 93117, USA}
\author{R. Graff}
\affiliation{Google Inc., Santa Barbara, California 93117, USA}
\author{E. Jeffrey}
\affiliation{Google Inc., Santa Barbara, California 93117, USA}
\author{J. Kelly}
\affiliation{Google Inc., Santa Barbara, California 93117, USA}
\author{E. Lucero}
\affiliation{Google Inc., Santa Barbara, California 93117, USA}
\author{J. Y. Mutus}
\affiliation{Google Inc., Santa Barbara, California 93117, USA}
\author{M. Neeley}
\affiliation{Google Inc., Santa Barbara, California 93117, USA}
\author{C. Neill}
\affiliation{Department of Physics, University of California, Santa Barbara, California 93106, USA}
\author{P. J. J. O'Malley}
\affiliation{Department of Physics, University of California, Santa Barbara, California 93106, USA}
\author{P. Roushan}
\affiliation{Google Inc., Santa Barbara, California 93117, USA}
\author{A. Shabani}
\affiliation{Google Inc., Venice, CA 90291, USA}
\author{V. N. Smelyanskiy}
\affiliation{Google Inc., Venice, CA 90291, USA}
\author{A. Vainsencher}
\affiliation{Google Inc., Santa Barbara, California 93117, USA}
\author{J. Wenner}
\affiliation{Department of Physics, University of California, Santa Barbara, California 93106, USA}
\author{H. Neven}
\affiliation{Google Inc., Venice, CA 90291, USA}
\author{John M. Martinis}
\affiliation{Department of Physics, University of California, Santa Barbara, California 93106, USA}
\affiliation{Google Inc., Santa Barbara, California 93117, USA}
\date{\today}


\maketitle

\section{Quasistatic flux noise measurement}
We refer to noise slower than the experimental repetition rate as quasistatic. To measure the quasistatic flux noise in $\Phi^x_t$ (i.e., noise in $\varepsilon$), we use a pulse sequence similar to that used in Ref. \cite{Lanting2009} combined with the signal processing techniques used in Refs. \cite{Sank2012, Yan2012}. This allows us to obtain data both well below and well above $1\,\rm{Hz}$, the latter being achieved by directly processing a binary sequence of single-shot measurements.

\begin{figure}[h!]
\begin{centering}
\includegraphics[width=.8\linewidth]{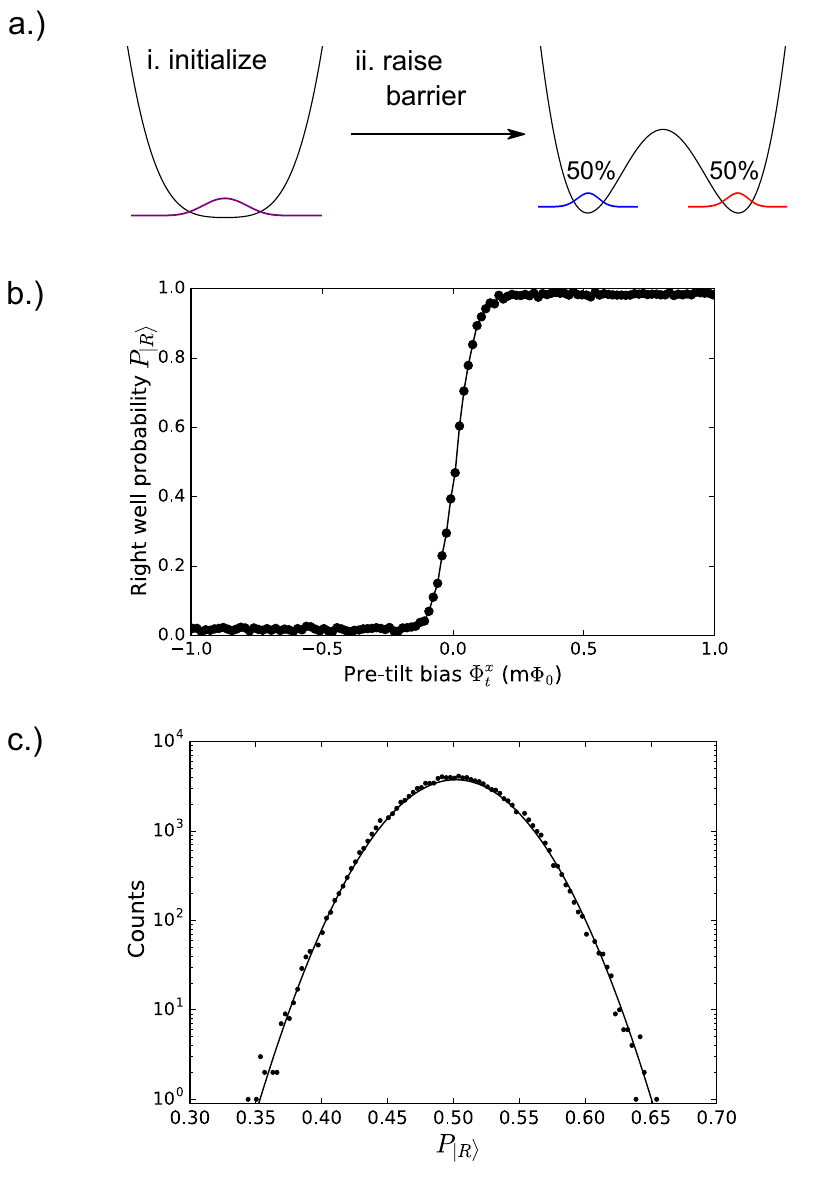}
\par\end{centering}
\caption{\footnotesize (a) Illustration of pulse sequence for the quasistatic flux noise measurement. (b) Calibration curve giving probability to end up in the right well as a function of external tilt flux bias. (c) Semi-log histogram of probabilities (obtained by averaging every 100 consecutive single-shot measurements) showing a Gaussian distribution with a standard deviation small enough to be within the linear part of the flux-probability curve.}
\label{figure:quasistatic_flux_noise}
\end{figure}

The measurement works as follows. We treat each experimental repetition as if it had a static flux offset, and repeatedly measure a function $f(\Phi^x_t(t))$ that is sensitive to fluctuations in $\Phi^x_t$ but not to fluctuations in $\Phi^x_{\rm{ba}}$. We do this by performing the pulse sequence illustrated in Fig. \ref{figure:quasistatic_flux_noise}(a). We initialize the qubit in its ground state (by energy relaxation) in the single-well regime at zero tilt. We then symmetrically raise the barrier so that in the absence of noise in $\Phi^x_t$ there would be probability 0.5 to end up in the left or right well, completely uncorrelated with any previous or future measurement. Deviations from $P=0.5$ correspond to deviations away from zero tilt in $\Phi^x_t$. We calibrate this experiment by measuring $P_{|R\rangle}$ (the probability of ending up in the right well) as a function of applied tilt bias $\Phi^x_t$, as shown in Fig. \ref{figure:quasistatic_flux_noise}(b). Since we accurately know the applied flux, this curve gives a direct calibration $P_{|R\rangle}(\Phi^x_t)$ between physical flux and probability, which (as detailed below) can be used to convert between probability noise and flux noise as long as the flux excursions are small enough that they remain on the linear part of the curve (close to $P_{|R\rangle}=0.5$). The exact functional form of this curve depends on the ramp rate due to factors including non-adiabatic transition and thermalization, but it is possible to choose a ramp rate such that the flux fluctuations remain within the linear part of the curve.
We verify this by looking at the raw measurement statistics of the noise measurement when nominally parked at $P_{\ket{R}} = 0.5$. For example, Fig. \ref{figure:quasistatic_flux_noise}(c) gives a histogram of probabilities obtained by averaging every 100 consecutive single-shot samples, along with a Gaussian fit. We also checked experimentally that the quasistatic sensitivity to the barrier bias, $dP_{\ket{R}}/d\Phi^x_{\rm{ba}}$, was less than $\frac{1}{100}\cdot dP_{|R\rangle}/d\Phi^x_t$, which is certainly negligible upon taking the square when comparing the relative contributions of incoherent flux noise.

For flux noise data below $\sim 0.1\,\rm{Hz}$, we measure a time series of $P_{\ket{R}}$ (each value being the average of a few hundred consecutive stats) over a total period of $\sim 24$ hours and use the same signal processing techniques as in Ref. \cite{Sank2012} to obtain $S^+_{\Phi_t^x}(f)$. For the data above $\sim 0.1 \,\rm{Hz}$, we instead obtain $S^+_{\Phi_t^x}(f)$ by processing time series of $N\approx 1$ million single-shot measurements taken with a regularly spaced sampling interval $\delta t$ (ranging from $10-100\,\mu$s) and use a variant of the signal processing techniques used in Ref. \cite{Yan2012}. We refer to this method as a ``1-bit detector measurement'' because it involves keeping all single-shot measurement results without explicitly computing probabilities. Given an underlying sequence of flux fluctuations $\{\delta \Phi_n\}$ in the qubit, we obtain a finite probabilistic binary sequence $\{x_n\}$ of length $N$, with $x_n\in \{-1, +1\}$ corresponding to each single-shot measurement, where $-1$/$+1$ are assigned to the outcomes $|L\rangle$/$|R\rangle$ respectively. The probability $P_{x_n}$ of obtaining $-1$ or $+1$ is related to the underlying sequence $\{\delta \Phi_n\}$ according to
\begin{align}
P_{x_n}(x) =&
\,\delta (x - 1) \left[\frac{1}{2} + \frac{dP}{d\Phi^x_t} \delta \Phi_n\right] \nonumber \\
+&\, \delta (x + 1) \left[\frac{1}{2} - \frac{dP}{d\Phi^x_t} \delta \Phi_n\right], \label{eq:P_x_n2}
\end{align}
where for short we now use $P$ to denote $P_{|R\rangle}$ and we have assumed a linear probability-flux curve. Defining the DFT coefficients as
\begin{equation}
\tilde{x}_k \equiv \sum_{n=0}^{N-1} x_n e^{-i 2 \pi n k / N},\label{dft_coefficients}
\end{equation}
then the expected value for the periodogram (PSD estimate) of $\{x_n\}$ can be related to the underlying power spectral density of $\{\delta\Phi_n\}$ according to
\begin{align}
\langle |\tilde{x}_k|^2 \rangle
&= \sum_{n,m = 0}^{N-1} \langle x_n x_m \rangle e^{-i 2 \pi (n-m) k / N}\nonumber\\
&=\sum_{n,m=0}^{N-1} \left[4\left(\frac{dP}{d\Phi^x_t}\right)^2 \delta \Phi_n \delta \Phi_m + \delta_{n,m} \right] e^{-i 2 \pi (n-m)\frac{k}{N}} \nonumber \\
&= 4\left(\frac{dP}{d\Phi^x_t}\right)^2\left| \sum_{n=0}^{N-1} \delta \Phi_n e^{-i 2 \pi n k / N} \right|^2 + N \nonumber \\
&= 4\left(\frac{dP}{d\Phi^x_t}\right)^2\left| \widetilde{\delta \Phi}_k \right|^2 + N,\label{eq:psd_floor}
\end{align}
where in the second line we have used the relation for the correlation
$$
\langle x_n x_m \rangle = \left\{
        \begin{array}{ll}
            4\left(\frac{dP}{d\Phi^x_t}\right)^2\delta\Phi_n\delta\Phi_m & \quad n\ne m \\
            1 & \quad n=m
        \end{array}
    \right.
$$
computed from (\ref{eq:P_x_n2}). We can then convert to a physical single-sided flux noise PSD after assigning a sampling interval time $\delta t$ to the sequence according to
\begin{align}
S^+_{\Phi}(f) &= \frac{2T}{N^2}\left| \widetilde{\delta \Phi}_{k=fT} \right|^2\nonumber\\
&= \frac{2 \delta t}{N} \frac{1}{4\left(\frac{dP}{d\Phi^x_t}\right)^2}\langle \left| \tilde{x}_{k=fT} \right|^2 \rangle - 2 \frac{1}{4\left(\frac{dP}{d\Phi^x_t}\right)^2}\delta t \, .\label{eq:physical_psd}
\end{align}
where $T = N\delta t$ is the total acquisition time for the dataset of $N$ samples. The normalization convention for $S^+_\Phi$ is chosen so that the total power in the signal is obtained by integrating over positive frequencies only. It is normalized such that the deterministic signal $\Phi(t) = A\sin(2\pi f t)$, with $f$ in the baseband of the sampled DFT, has total power $A^2/2$ as physically expected. The ability to correctly extract the underlying $S^+_{\Phi}(f)$ with these formulas was verified by numerically feeding in an artificially generated noise source into a numerical simulation of our measurement, which was in turn fed into the data analysis software.

Equation (\ref{eq:physical_psd}) tells us that the PSD of our measurement sequence $S^+_x(f)$ will be a combination of the underlying $S^+_\Phi(f)$ plus a white noise floor. This noise floor makes sense because if there were no flux noise at all, we would expect shot noise from a perfectly uncorrelated probability of $0.5$ for each measurement result. This white noise floor can be substantial: for typical parameters of $\delta t = 10-100\,\mu$s and $dP/d\Phi^x_t \approx 2000/\Phi_0$, it has an equivalent flux noise amplitude of order $1\,\mu \Phi_0/\sqrt{\rm{Hz}}$. Since this is the typical strength expected for the intrinsic $1/f$ flux noise of the device, this means the white noise floor will dominate the signal above $\sim1\,\rm{Hz}$.

Fortunately, there is a way to process the data that allows one to drastically reduce the shot noise floor and infer $S^+_\Phi(f)$ from the measurement of $S^+_x(f)$ without noticeably distorting the underlying flux noise signal \cite{Yan2012}. The idea is inspired by the technique of using two separate detectors to sample a signal and computing the cross spectrum to throw away the detectors' contribution to the noise. By breaking $\{x_n\}$ into two interleaved series and taking the cross-spectral density (CSD) of the two sub-series, in the limit of infinite statistics this would completely eliminate the white noise floor, as shown below. The intuition behind this is that the interleaving removes the zero-delay autocorrelation term because the point $\delta\Phi_n$ never ``sees itself'' in the sum. However, the actual datasets are finite, so as discussed below this cancellation will not be perfect. Regarding the underlying $1/f$ signal itself, at least at low enough frequencies, the two interleaved flux noise signals should be approximately equal because the noise is highly correlated, and so the spectrum should not become distorted. This will be shown numerically below.

Mathematically, we define the two subsequences
\begin{equation}
x_n' = x_{2n},\,\,\, x_n'' = x_{2n + 1}
\end{equation}
$n = 0, 1, 2, ... , M-1$, where $M=N/2$. We define a CSD for these interleaved sequences multiplied by a particular frequency-dependent phase factor:
\begin{align}
&\left\langle \tilde{x}_k'\left(\tilde{x}_k''\right)^*\right\rangle e^{i 2 \pi k /N}\label{eq:phase_shifted_CSD}
\\&\,\,\,\,\,\,\,\,\,\,= \sum_{n,m = 0}^{M-1} \langle x_n' x_m'' \rangle e^{i 2 \pi (n-m) k / M}e^{i 2 \pi k /2 M}\nonumber\\
&\,\,\,\,\,\,\,\,\,\,= 4\left(\frac{dP}{d\Phi^x_t}\right)^2 \sum_{n,m = 0}^{M-1} \delta\Phi_{2n}\delta\Phi_{2m+1}e^{-i 2 \pi (n-m-\frac{1}{2}) k / M},\label{eq:interleave_floor}
\end{align}
where the extra phase factor in line (\ref{eq:phase_shifted_CSD}) corrects for the phase shift arising from the time domain offset between the two interleaved sequences. This phase offset ensures that the interleaved CSD of a deterministic sinusoidal signal with integer wave index $q$ is itself real.

\begin{figure}[h!]
\begin{centering}
\includegraphics[width=1\linewidth]{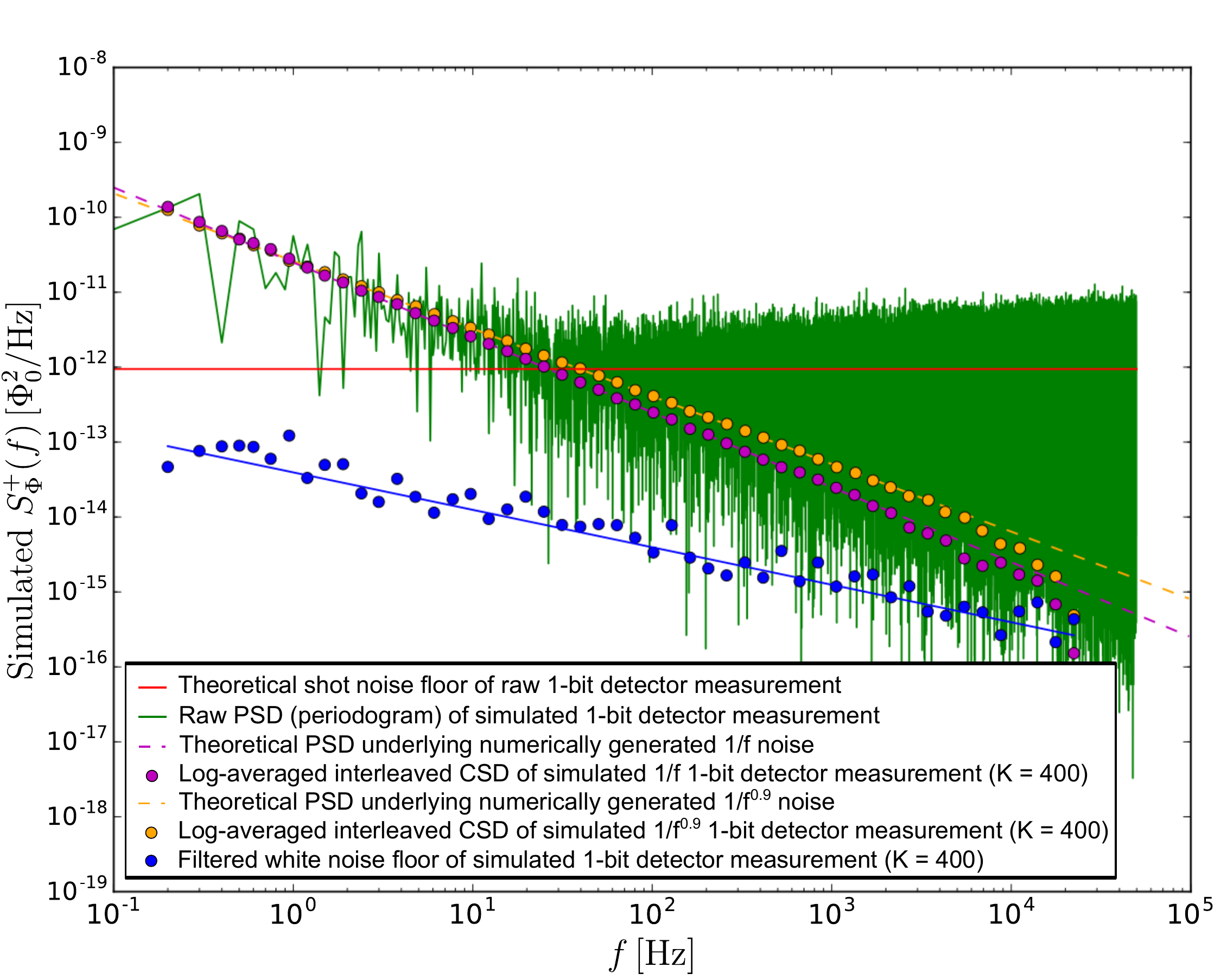}
\par\end{centering}
\caption{\footnotesize Numerical simulation of the 1-bit detector flux noise measurement for artificially generated $1/f$ and $1/f^{0.9}$ noise processes. The red line shows the expected white noise level for the raw 1-bit detector measurement without interleaving. Green shows the raw PSD of the simulated 1-bit detector measurement for a single realization of the numerically generated $1/f$ noise. The dashed magenta line shows the theoretical PSD used to generate the artificial $1/f$ noise. Magenta circles show the result of the simulated 1-bit detector measurement using the same processing that was used on the experimental data (interleaved + log-averaged over frequency, averaged over $K=400$ realizations of the numerically generated $1/f$ noise). Orange: same but for numerically generated $1/f^{0.9}$ noise. The blue circles show the white noise floor of the simulated measurement after interleaving and averaging, with the blue line its theoretically expected level.}
\label{figure:sim}
\end{figure}

Equation (\ref{eq:interleave_floor}) says that if we perform the interleave and take the CSD in the absence of underlying flux noise, we would observe absolutely no noise (since $\delta \Phi_n$ is identically zero). Of course, the only reason the noise cancellation is exact is due to the use of the expectation value operator, $\langle \cdot \rangle$, which means an average over an ensemble of infinitely many realizations of $\{x_n\}$ generated by a given $\{\delta\Phi_n\}$. In reality, we deal with finite sequences ($N \approx 10^6$), and instead of (\ref{eq:phase_shifted_CSD}), we can only compute the CSD coefficients $\tilde{x}_k'\left(\tilde{x}_k''\right)^* e^{i 2 \pi k /N}$ for a finite number of finite sequences. The simplest way to understand the effect of interleaving on the white noise floor in the actual experiment, then, is to think about a random walk of phasors in the complex plane. First, let us consider how we can reduce the noise floor within a single dataset, and then we will consider averaging over multiple datasets. When $\{\delta \Phi_n\}$ is identically zero (no flux noise), each CSD coefficient of the 1-bit detector measurement will itself a Gaussian distributed complex random variable, with uniformly distributed phase and some distribution of magnitude that is peaked close to $\delta t/(dP/d\Phi_t^x)^2$. Thus, the white noise floor will not actually be reduced unless we perform some sort of averaging, either across frequency bins or across datasets. Since it is informative to plot flux noise on a log-log scale, it is natural to use a logarithmic averaging scheme where the number of neighboring frequency bins whose CSD coefficients are averaged together is proportional to frequency. In other words, the number of bins per decade of frequency used in the averaging is constant, as is done for the quasistatic flux noise data shown in the main paper. Since the number of points $N$ averaged in a frequency bin is chosen to scale as $f$, and sum of $N$ random phasors scales in magnitude as $\sqrt{N}$, then upon taking the average over each frequency bin we would expect the white noise floor power to decrease as $1/\sqrt{N}$, meaning as $1/\sqrt{f}$. Thus, the act of interleaving and taking a coherent logarithmic frequency average within a given dataset \emph{amounts to a $1/\sqrt{f}$ filter for the white noise floor}. This can be seen from the slope of the observed filtered white noise floor in the numerical simulation of Fig. \ref{figure:sim}.

However, we find that the filtered suppression of the white noise floor from averaging over frequency bins within a single dataset alone is not enough to bring the white noise floor below the qubit flux noise signal at high frequencies. Because of this, we further average the CSD coherently over $K$ datasets, with $K$ a few hundred, before taking the real part. This gives a factor of $\sqrt{K}\approx 10$ further reduction in the filtered white noise floor power without distorting the underlying (presumably) correlated flux noise signal, according to the numerical simulation of the measurement shown in Fig. \ref{figure:sim}. This simulation uses artificially generated $1/f^\alpha$ noise signals with $\alpha=1$ and $\alpha=0.9$ and magnitude at $1\,\rm{Hz}$ equal to $5\,\mu\Phi_0/\sqrt{\rm{Hz}}$, the value extracted from experiment. We note that the discretely generated artificial $1/f$ noise extends an order of magnitude higher in frequency than the sampling frequency of the simulated 1-bit detector measurement. This was done purposefully to make sure there is no influence from aliasing. We find that the interleaving technique greatly reduces the effects of aliasing that would otherwise be present from the substantial amount of $1/f$ noise that is likely present above the Nyquist frequency of the measurement. From the simulation, we see that the 1-bit detector measurement after interleaving and coherently averaging over datasets would faithfully reproduce the $1/f$ power spectrum without any significant distortions except for the highest half-decade of frequency. The filtered white noise is well below the simulated $1/f$ noise (below the highest half-decade of frequency), which means it is even further below the experimentally measured flux noise (because of the significant ``bump'' observed starting at $10 - 100\,\rm{Hz}$ in the experimental data) \footnote{We note that the ``bump'' in noise around $100\,\rm{Hz}$ in the experimental data is unchanged when the pulse tube compressor of the dry dilution refrigerator is turned off.}. As an extra consistency check, we note that the extracted flux noise data in the experiment was not materially changed (apart from the highest factor of 2 in frequency) whether we coherently averaged over 250 or 500 datasets.

\section{Consistent definition of $S_\Phi^+(f)$ at low and high frequencies}
Because the low and high frequency flux noise are measured by very different methods, we must be careful to have consistent definitions of $S^+_\Phi(f)$ at low and high frequencies. In other words, in Fig. 2(e) of the main text, we must be sure we are plotting the same physical quantity at low and high frequencies, without any discrepant factors of 2 or 2$\pi$. Such a discrepancy could for example affect the best-fit value of $\alpha$ in an interpolating power law between the two frequency ranges.

To infer $S^+_\Phi(f)$ at low frequencies, we measure the discrete time sequence $\Phi_n$, where $\Phi_n$ is a classical real number, over $N$ discrete time steps indexed by $n\in\{0, 1, ... , N-1\}$ and separated by the physical sampling interval $\delta t$ (meaning total data acquisition time $T = (N-1)\delta t \approx N\delta t$). We then estimate the single-sided PSD by computing
\begin{equation}
S^+_\Phi(f) = \frac{2 T}{N^2} \langle |\tilde{\Phi}_{k=fT}|^2\rangle,\label{low_freq_def}
\end{equation}
where the DFT coefficients $\Phi_k$ are defined by
\begin{equation}
\tilde{\Phi}_k \equiv \sum_{n=0}^{N-1} \Phi_n e^{-i 2 \pi n k / N}.\label{dft_coefficient}
\end{equation}
and $\langle \cdot \rangle$ denotes an ensemble average since $\tilde{\Phi}_k$ is itself a random variable.

At high frequencies, we instead infer the flux noise spectral density through Fermi's golden rule \cite{Schoelkopf2003} using the fact that $1/T_1 = \Gamma_\uparrow + \Gamma_\downarrow$, which implies the relation
\begin{align}
S_\Phi^+(f) &\equiv S_\Phi(f) + S_\Phi(-f)\nonumber \\
&= \frac{\hbar^2}{T_1}\frac{1}{|\langle 0|\frac{d\hat{H}}{d\Phi^x_t}|1\rangle|^2} = \frac{\hbar^2 L^2}{T_1}\frac{1}{|\langle 0|\hat{\Phi}|1\rangle|^2},\label{high_freq_def}
\end{align}
where \cite{Schoelkopf2003}
\begin{equation}
S_\Phi(\omega) \equiv \int_{-\infty}^\infty d\tau e^{i\omega \tau} \langle \Phi(\tau ) \Phi(0)\rangle.\label{schoelkopf_definition}
\end{equation}
Here, $\Phi$ could be an operator, but for simplicity we can assume it's a real number, since it is sufficient to check whether or not the two definitions (\ref{low_freq_def}) and (\ref{high_freq_def}) for $S^+_\Phi$ coincide for a classical incoherent flux noise source acting on the qubit. Showing this will turn out to be equivalent to deriving the Wiener-Khinchin theorem for a stationary stochastic process.

First, we can write (\ref{dft_coefficient}) in the continuum limit $N\to\infty, \delta t\to 0$ with $T$ held constant, so that $\Phi(t=n\delta t) = \Phi_n$ and $\tilde{\Phi}_{k=ft} \to \frac{1}{\delta t} \int_0^T \Phi(t) e^{-i2\pi f t}dt$. Since $\Phi(t)$ is a random process, so is $\tilde{\Phi}(f)$, so we keep the $\langle \cdot \rangle$ before taking the limit $T\to\infty$. We can obtain its expectation value by taking the limit $T\to \infty$ after plugging the continuum limit expression into (\ref{low_freq_def}),
\begin{align}
S^+_\Phi(\omega) &= \lim_{T\to\infty}\frac{2}{T}\left\langle\left|\int_0^T \Phi(t)e^{-i \omega t} dt \right|^2\right\rangle\nonumber\\
&=\lim_{T\to\infty} \frac{2}{T}\int_0^T \int_0^T dt\,dt'\,e^{i\omega t} e^{-i\omega t'} \langle\Phi(t)\Phi(t')\rangle\nonumber\\
&=\lim_{T\to\infty} \frac{2}{T}\int_0^T \int_0^T dt\,dt'\,e^{-i\omega (t-t')} \langle\Phi(t-t')\Phi(0)\rangle,\label{tricky_integral}
\end{align}
where in the last step we have assumed that $\Phi(t)$ is a stationary process. To continue, we note that the integrand [call it $f(t, t')$] inside the double integral is a function of $\tau \equiv t' - t$ alone, meaning that $f(\tau) = f(t - t')$ is constant along lines of constant $\tau$ defined by the equation $t' = t + \tau$ within the $t$-$t'$ plane. We can therefore convert the double integral into a one-dimensional integral in $f(\tau)$, by integrating the diagonal ``slices'' formed by such lines across the two-dimensional domain of integration. The domain of integration is the square $[0,T] \times [0,T]$ in the $t$-$t'$ plane, which is covered by diagonal strips parameterized by $\tau$ ranging from $-T$ to $T$. The area of each diagonal strip corresponding to $\tau$ with infinitesimal width $d\tau$ is $\sqrt{2}(T - |\tau|)\frac{d\tau}{\sqrt{2}} = (T - |\tau|)d\tau$, so we can convert $\int_0^T \int_0^T dt\,dt'\,f(t - t')$ to $\int_{-T}^{T}d\tau\,f(\tau)$, meaning that (\ref{tricky_integral}) becomes
\begin{align}
S^+_\Phi(\omega) &= \lim_{T\to\infty} \frac{2}{T} \int_{-T}^T d\tau \, e^{i\omega \tau}\langle \Phi(\tau)\Phi(0)\rangle (T - |\tau|)\nonumber\\
&= \lim_{T\to\infty} 2 \int_{-T}^T d\tau \, e^{i\omega \tau}\langle \Phi(-\tau)\Phi(0)\rangle \left(1 - \frac{|\tau|}{T}\right)\nonumber\\
&= 2\int_{-\infty}^\infty d\tau \, e^{i\omega \tau} \langle \Phi(\tau)\Phi(0)\rangle,
\end{align}
where in the last line we have used the property that the autocorrelation function $\langle x(t+\tau)x(t)\rangle$ is an even function of $\tau$. Comparing this to (\ref{schoelkopf_definition}) and (\ref{high_freq_def}) shows that the two definitions of $S^+_\Phi$ are indeed equivalent.

\section{Flux noise at high and low frequencies changes similarly between samples}
\begin{figure}[h!]
\begin{centering}
\includegraphics[width=1.0\linewidth]{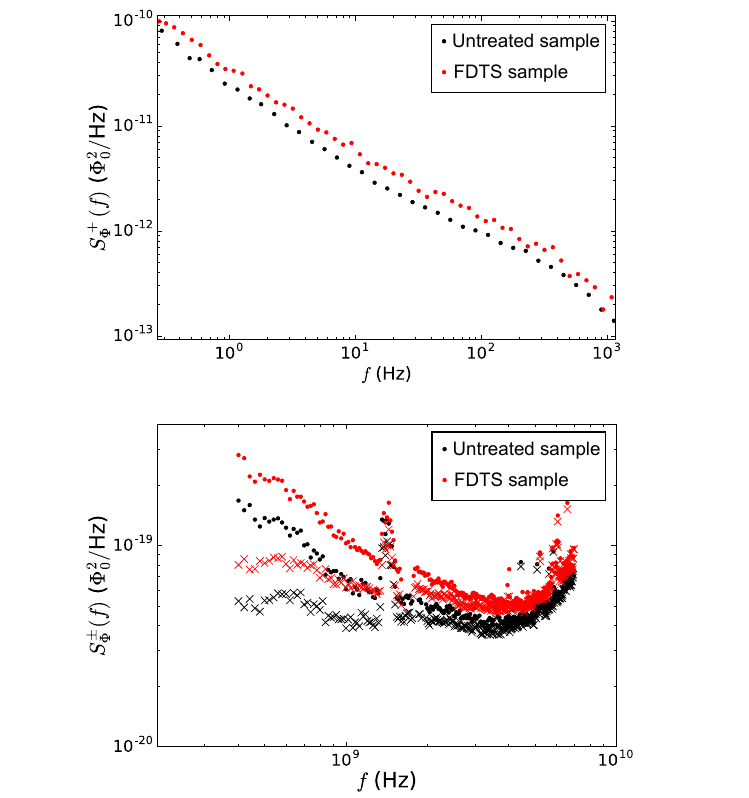}
\par\end{centering}
\caption{\footnotesize Comparison of flux noise at low and high frequencies between two different samples at base temperature. The samples were nominally identical apart from differing fabrication post-treatments.}
\label{figure:low_and_high}
\end{figure}

Fig. \ref{figure:low_and_high} shows low and high\footnote{Regarding the ``mode'' at $1.4\,\rm{GHz}$, this mysterious feature is present at the same frequency and similar strength in all qubits with different chip and sample box sizes, and differing types of filtering and attenuation on the bias lines, including the addition of attenuation on the output of the readout line.} frequency flux noise data for nominally the same qubit on two different chips, with the second chip seeing extra post-processing in the form of a downstream oxygen ash clean and the application of a (nominal) monolayer of perfluorodecyltrichlorosilane (FDTS) via molecular vapor deposition. The second sample also sat covered in photoresist for 6 months longer than the first sample. The flux noise power just below 1 GHz changes by a factor of 1.6, while the flux noise power around 1 Hz changes by a factor of approximately 1.5. Consistent with this we also observe that the Ramsey decay time $T_{\phi 2}$ away from zero tilt was $\sim30\%$ lower on the FDTS sample when measured at a point with the same sensitivity of $f_{10}$ to tilt flux (this degradation was reproducible at several bias points).


\section{Checking for distortion of extracted $S^\pm_{\Phi^x_t}(f)$ from nonlinear crosstalk}
In the main text, to deduce the high frequency flux noise $S^\pm_{\Phi^x_t}(f)$ in the main qubit loop we measure $T_1$ at zero tilt, because at degeneracy only flux noise in $\varepsilon(\Phi^x_t)$ and not noise in $\Delta(\Phi^x_{\rm{ba}})$ would induce transitions between qubit energy eigenstates. However, this is no longer strictly true if the two junctions in the DC SQUID are not perfectly symmetric, which is to be expected due to fabrication imperfections. It can be shown \cite{Koch2007Transmon} that if the junction asymmetry is $d\equiv \frac{E_{J1} - E_{J2}}{E_{J1} + E_{J2}}$, any flux threading the DC SQUID loop will lead to an offset in tilt flux according to the nonlinear relation
\begin{equation}
\Delta \Phi^x_t = \frac{\Phi_0}{2\pi}\tan^{-1}\left(d \tan\left[\pi \Phi^x_{\rm{ba}}/\Phi_0\right]\right),
\end{equation}
implying that noise in $\Phi^x_{\rm{ba}}$ leads to noise in $\Phi^x_t$ according to the differential transfer function
\begin{equation}
\frac{d\Phi^x_t}{d\Phi^x_{\rm{ba}}} = \frac{d}{2}\frac{1}{d^2 \sin^2(\pi \Phi^x_{\rm{ba}}/\Phi_0) + \cos^2(\pi \Phi^x_{\rm{ba}}/\Phi_0)}.
\end{equation}

By measuring $\Delta \Phi^x_t$ as a function of $\Phi^x_{\rm{ba}}$ and subtracting out any contribution from linear geometric crosstalk between barrier and tilt bias lines, we can extract this intrinsic nonlinear transfer function experimentally. For the data in the main text, we obtain $(\frac{d\Phi_t}{d\Phi_{\rm{ba}}})^2 < .005$ over the range of $\Delta(\Phi^x_{\rm{ba}})$ measured, corresponding to a junction asymmetry of $\sim1\%$. This suggests that dissipation from incoherent flux noise is likely dominated by noise in $\Phi^x_t$ and not $\Phi^x_{\rm{ba}}$, since the $1/f$ noise in $\Phi^x_{\rm{ba}}$ should be less than or comparable to the noise in $\Phi^x_t$ (using conventional Ramsey experiments \cite{Sank2012} at zero tilt we obtain the noise in the DC SQUID flux $\Phi^x_{\rm{ba}}$ has magnitude $1 - 2\,\mu\Phi_0/\sqrt{\rm{Hz}}$ at $1\,\rm{Hz}$, compared to $\sim 5$ $\mu\Phi_0/\sqrt{\rm{Hz}}$ for $\Phi^x_t$). However, this does not exclude the possibility of the relative strength of noise in $\Phi^x_t$ and $\Phi^x_{\rm{ba}}$ changing greatly between $1\,\rm{Hz}$ and $\,\rm{GHz}$, but given that we see similar $T_1$'s over several samples with different junction asymmetries $d$, this seems not to be the case. Note that this analysis does not rule out noise from surface spins fluctuating on the wiring of the DC SQUID, because this wire is shared by both barrier and tilt loops; rather, it only implies that noise from such fluctuators would affect the $T_1$ data via induced noise in $\Phi^x_t$ and not due to induced noise in $\Phi^x_{\rm{ba}}$.

\section{Checking for dissipation from non-equilibrium quasiparticles}
The frequency dependence of the fluxmon $T_1$ at zero tilt below $\sim3\,\rm{GHz}$ is also consistent with quasiparticle dissipation if one were to assume a large enough population of non-equilibrium quasiparticles in the system. We give theoretical and experimental arguments as to why this is unlikely to be a dominant effect in our system, including a test of the effects of magnetic vortices.

Quasiparticles with energy near the superconducting gap $\Delta$ can absorb energy from the qubit when they tunnel across one of the Josephson junctions, while ``hot'' quasiparticles with energies more than $\hbar \omega_{10}$ from the gap can excite the qubit. If the energies of all quasiparticles influencing the qubit are sufficiently less than $2\Delta$, then for an arbitrary quasiparticle occupation distribution $f(E)$, the decay and excitation rates induced on the qubit are given by \cite{Catelani2011}
\begin{equation}
\Gamma_{i\to f} = \sum_{j=1,2}\left|\langle f|\sin \frac{\hat{\phi}_j}{2}|i\rangle\right|^2 S_{\rm{qp}}^j(\omega_{if}),\label{qp_rate}
\end{equation}
where $j$ indexes the two Josephson junctions of the fluxmon and
\begin{align}
S_{\rm{qp}}^j(\omega) = \frac{32 E_{Jj}}{\pi\hbar}\int_0^\infty &dx\, \rho((1+x)\Delta)\rho((1 + x)\Delta + \hbar\omega)\times\nonumber\\
&f[(1 + x)\Delta](1 -f[(1+x)\Delta+\hbar\omega])\label{S_qp}
\end{align}
is the double-sided quasiparticle current spectral density, with $\rho(E) = E/\sqrt{E^2 - \Delta^2}\approx 1/\sqrt{2(E - \Delta)/\Delta}$ the normalized quasiparticle density of states. The formula (\ref{S_qp}) works for $\omega < 0$ by simply swapping the arguments of $f$ and replacing $\omega$ with $-\omega$. In a non-tunable gap flux qubit (i.e., a single-junction fluxmon), at zero tilt the junction would be biased at $\pi$, meaning that quasiparticle dissipation ($\propto |\langle 0|\sin \frac{\hat{\phi}}{2}|1\rangle|^2$) would vanish (physically, this is due to destructive interference between electron-like and hole-like tunneling \cite{Pop2014}). However, for the gap-tunable fluxmon, even though the effective dynamical phase $\hat{\phi} = (\hat{\phi}_1 + \hat{\phi}_2)/2$ can be biased at $\pi$ at zero tilt, the phase of the individual junctions are not. Instead, flux quantization dictates that $\hat{\phi}_1 = \hat{\phi} - \pi \Phi_{\rm{ba}}/\Phi_0, \hat{\phi}_2 = \hat{\phi} + \pi \Phi_{\rm{ba}}/\Phi_0$, meaning that the matrix element can be non-zero and quasiparticle dissipation can occur even at zero tilt.

While a thermal distribution of quasiparticles would be much too small to explain the observed dissipation, this does not rule out the possibility of non-equilibrium quasiparticles. A reasonable model for computing the distribution of non-equilibrium quasiparticles is outlined in Refs. \cite{Martinis2009, Lenander2011, Wenner2013}. Here, quasiparticles are assumed to be injected at some energy or range of energies well above the gap and, via electron-phonon scattering and recombination, relax to some steady-state distribution that is essentially independent of the injection energy as long as the injection energy is high enough. Using the steady state equations outlined in Ref. \cite{Lenander2011}, we can calculate the expected distribution of nonequilibrium quasiparticles as a function of injection rate and phonon temperature $T_{\rm{ph}}$, and then using (\ref{qp_rate}) and (\ref{S_qp}) numerically calculate the resulting up and down transition rates to obtain the quasiparticle-induced $T_1$ and stray population. We choose an injection rate that leads to a quasiparticle density that best matches the measured $T_1$ data while choosing the phonon temperature to be equal to the fridge temperature. We assume the same injection rate for all four fridge temperatures used in the main paper. The injection energies were between $2.1\Delta$ and $2.2\Delta$, though changing this energy range does not materially affect the result. The results are summarized in Fig. \ref{figure:qp_fig}, where we plot the computed $f(E)$, $T_1$ and stray population/effective temperature induced on the qubit on top of the $T_1$ and stray population data used in Fig. 3 of the main paper. The best fit yields a quasiparticle density $n_{\rm{qp}}\equiv 2D(E_F)\int_\Delta^\infty\rho(E)f(E)\,dE$ of $3.5/\mu{\rm{m}}^3$, or a fractional quasiparticle density $x_{\rm{qp}} = n_{\rm{qp}}/n_{cp}=n_{\rm{qp}}/(D(E_F)\Delta)=1.3\times 10^{-6}$, where $D(E_F)$ is the density of electron states at the Fermi energy.

\begin{figure}[h!]
\begin{centering}
\includegraphics[width=1.0\linewidth]{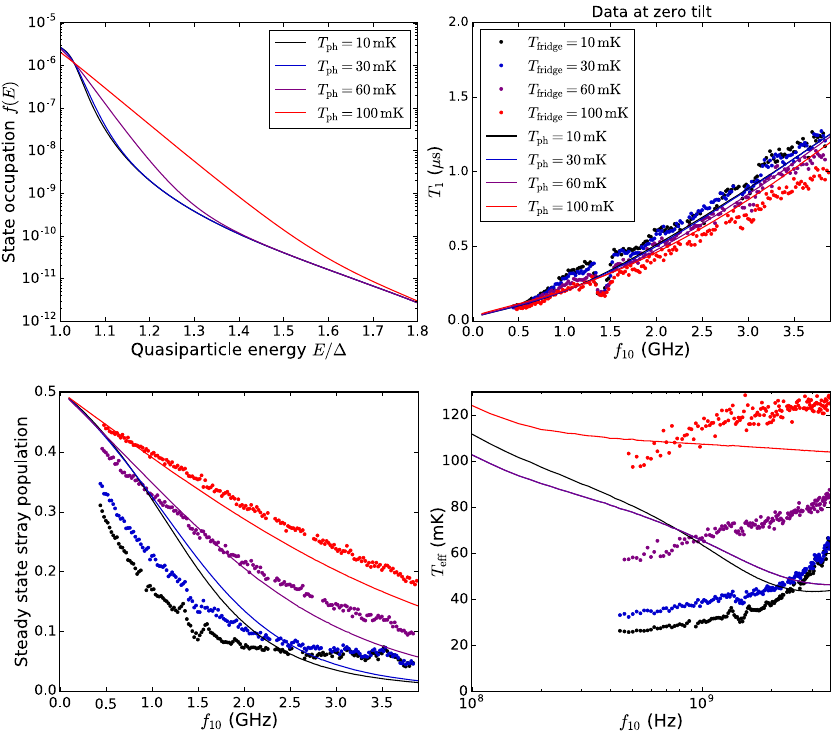}
\par\end{centering}
\caption{\footnotesize Data vs. fit to nonequilibrium quasiparticle theory. Although a quasiparticle density can be chosen large enough to roughly match $T_1$ vs. frequency, a simultaneous fit to stray population is not possible within this model.}
\label{figure:qp_fig}
\end{figure}

We make several observations about this calculation. While the fits do match the $T_1$ data reasonably well below $\sim 4\,\rm{GHz}$, the stray population is quite off and does not have nearly a large enough dependence on temperature at low frequencies. We also see that the quasiparticle-induced qubit effective temperature does not approach a constant value at low frequencies like the data appears to do, and like the stray populations, the predicted effective temperatures do not match the data well at all, especially for the lowest three temperatures. Furthermore, there are several experimentally-based estimates substantially below $x_{\rm{qp}}=1.3\times10^{-6}$ in the superconducting qubit literature, for example $1\times 10^{-8}$, $4\times 10^{-8}$, and $3\times10^{-7}$ in Refs. \cite{Riste2013, Vool2014, Wang2014}, and $2.2\times 10^{-7}$ in Ref. \cite{Jin2015} (where the nonequilibrium quasiparticle density explained both $T_1$ and stray population), so it is not unreasonable that $x_{\rm{qp}}$ is similarly low in our system, especially given the stringent level of light-tight filtering used in our setup \cite{Barends2011}. In addition, we note that if the non-equilibrium quasiparticle density was indeed $x_{\rm{qp}} = 1.3\times 10^{-6}$, the $T_1$ limit imposed on our standard Xmon qubits would be $7\,\mu\rm{s}$ at $3\,\rm{GHz}$, but Xmons consistently achieve $T_1$ a factor of 10 higher than this at $3\,\rm{GHz}$, and furthermore show a defect structure and opposite overall scaling in frequency from that predicted from quasiparticles. We have no reason to believe the quasiparticle density would be any higher in the fluxmon device given that the device is fabricated with identical materials and chip mount, and the fridge wiring is essentially identical to that used for Xmon experiments. If the quasiparticles were being introduced by the long (few $\mu\rm{s}$) readout pulses used for the fluxmon, we would expect $T_1$ to depend on the repetition time between experiments. We do not see any change in $T_1$ as we vary the repetition time from $50\,\mu\rm{s}$ to $1000\,\mu\rm{s}$.

One more experimental check we can perform is to see if $T_1$ improves after inducing magnetic vortices into the Al film of the sample. Abrikosov vortices, which have quasinormal cores, form when a thin-film superconductor is cooled through $T_c$ in the presence of a magnetic field. It is well-known that such vortices trap quasiparticles, and have even been shown to significantly decrease quasiparticle-induced dissipation in superconducting qubits \cite{Wang2014}. In particular, for a thin-film Al transmon qubit vortices were shown to significantly decrease quasiparticle-induced dissipation with a modest applied field of $\sim 10\,\rm{mG}$ \cite{Wang2014}. To check if there is any improvement in $T_1$ to be gained from the presence of vortices, we added a magnetic coil to the setup directly outside the qubit box, which we used to apply a several different magnetic fields between $-30$ and $30\,\rm{mG}$ to the qubit chip as it cooled through its superconducting transition. The result of these magnetic field cools are shown in Fig. \ref{figure:field_cools}. We see only degradation of qubit $T_1$ with applied magnetic field. This data suggests that quasiparticles are not playing a dominant role in qubit dissipation in the fluxmon device.

\begin{figure}[h!]
\begin{centering}
\includegraphics[width=.9\linewidth]{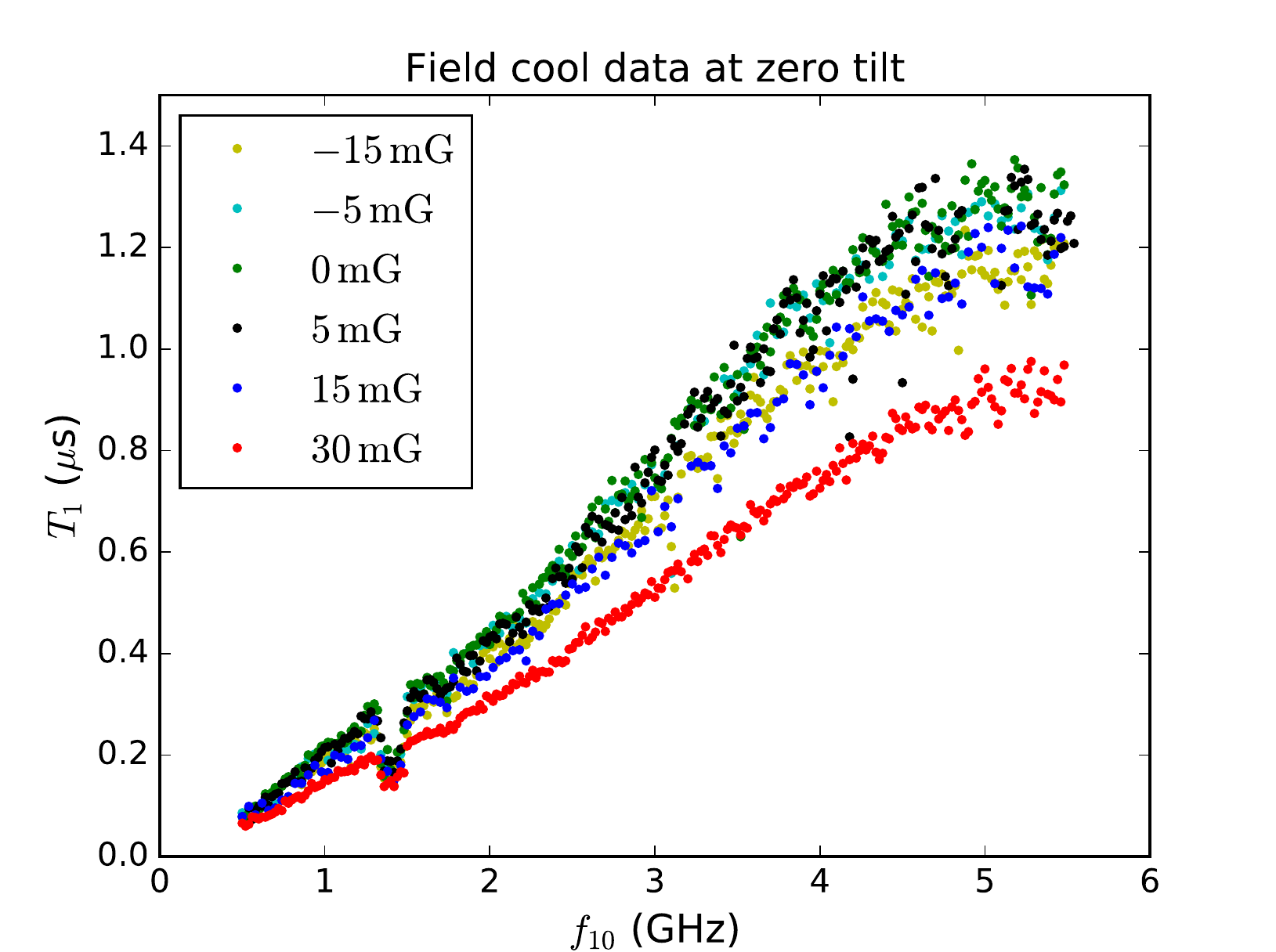}
\par\end{centering}
\caption{\footnotesize Field cool data. Inducing vortices only degrades $T_1$.}
\label{figure:time_traces}
\end{figure}

\begin{figure}[h!]
\begin{centering}
\includegraphics[width=.78\linewidth]{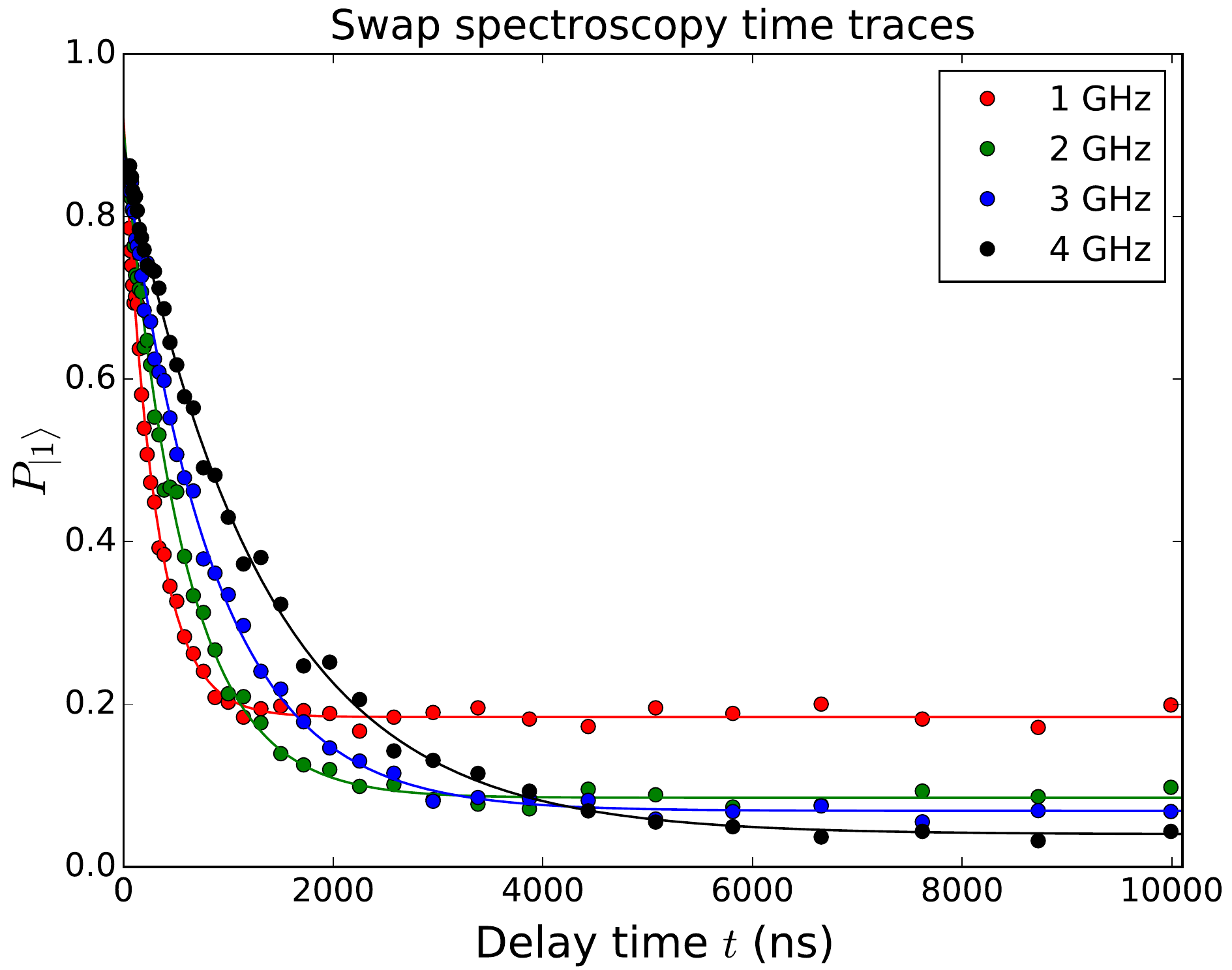}
\par\end{centering}
\caption{\footnotesize Swap spectroscopy time traces at zero applied field show an exponential time dependence. State preparation infidelity is due to imperfect $\pi$ pulse given the short $T_2$ time, and relaxation during the detuning pulse.}
\label{figure:field_cools}
\end{figure}

We also note that, as depicted in Fig. \ref{figure:time_traces} for the zero field data, the raw $T_1$ traces of 1-state population versus time fit very well to an exponential decay as opposed to showing a measurable non-exponential behavior that could be due to fluctuation in quasiparticle number \cite{Yan2016}.

\section{Low and high frequency noise cutoffs for Macroscopic Resonant Tunneling Rates}
In the main text, it was argued that the low-frequency cutoff $f_l$ in the noise integral for the MRT tunneling linewidth $W$ should be the inverse of the total experimental data acquisition time, whereas the $f_l$ for the integral for the reorganization energy $\varepsilon_p$ should be the tunneling rate near maximum tunneling. The latter physically makes sense as the reorganization energy should not depend on the time at which the experiment was performed. In other words, only dissipation at frequencies higher than the slowest timescale of the tunneling can affect this energy. However, different instances of the tunneling experiment may have different quasistatic flux offsets $\delta \varepsilon = 2 I_p \delta \Phi^x_t$ as discussed earlier in the context of the low frequency flux noise measurement, which can give additional broadening of the MRT tunneling peak. For simplicity, let us take the Gaussian approximation to the lineshape described in the main text, $\Gamma(\varepsilon) = \sqrt{\frac{\pi}{8}} \frac{\Delta^2}{\hbar W} \exp\left[-\frac{(\varepsilon - \varepsilon_p)^2}{2 W^2}\right]$, so that what is actually measured after averaging is the quantity
\begin{equation}
\overline{\Gamma}(\varepsilon) = \int_{-\infty}^\infty\, d(\delta \varepsilon) \Gamma(\varepsilon + \delta \varepsilon)p(\delta \varepsilon),
\end{equation}
where $p(\delta \varepsilon) = \frac{1}{\sqrt{2\pi\sigma^2}} e^{-\frac{(\delta \varepsilon)^2}{2 \sigma^2}}$ describes the Gaussian distribution of quasistatic flux fluctuations. Performing the integration yields
\begin{equation}
\overline{\Gamma}(\varepsilon) = \frac{\pi}{8}\frac{\Delta^2}{\sqrt{W_0^2 + \sigma^2}}e^{-\frac{(\varepsilon - \varepsilon_p)^2}{2\left(W_0^2 + \sigma^2\right)}},
\end{equation}
which shows that the original $W_0$ (obtained from integrating down to $f_l$ equal to the maximum tunneling rate) is broadened by the r.m.s. quasistatic flux fluctuations via addition in quadrature. Since $W^2 = 4 I_p^2 \int_{f_l}^{f_h} df\,S^+_\Phi(f)$, this amounts to extending $f_l$ down to the inverse total experimental averaging time.

For the high frequency cutoff to the tunneling rate integral, we use the oscillation frequency of the inverted potential barrier. Previous theoretical and experimental work in macroscopic quantum tunneling has shown this to be the physical high-frequency cutoff \cite{Buttiker1982, Leggett1984, Martinis1988, Esteve1989}. Nevertheless, we find that the tunneling rate near resonance according to full integral in Eq. (3) of the main text is not materially affected by this high frequency cutoff.

\section{Temperature independence of the classical low frequency flux noise}
\begin{figure}[h!]
\begin{centering}
\includegraphics[width=.85\linewidth]{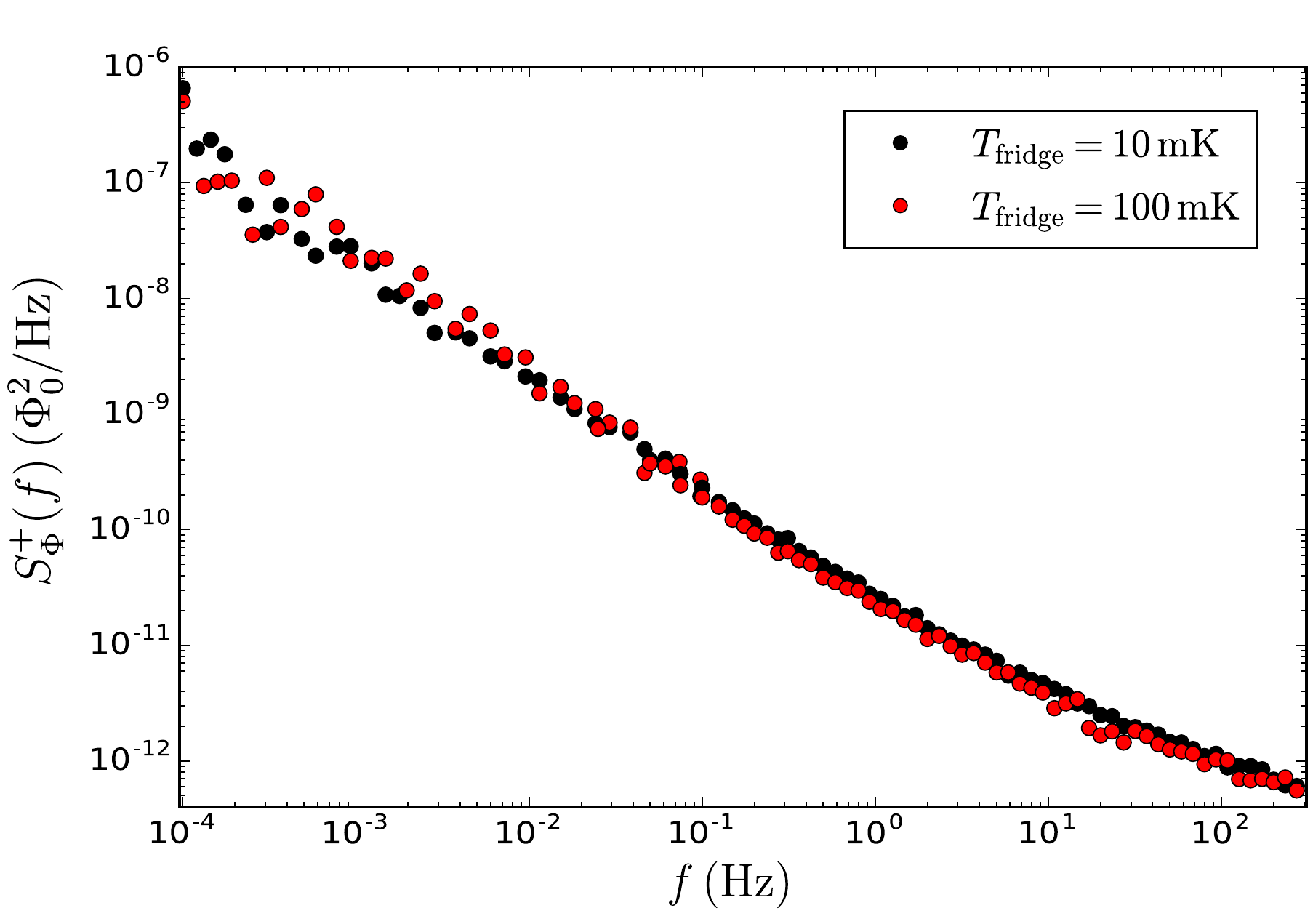}
\par\end{centering}
\caption{\footnotesize Quasistatic flux noise measurement is not noticeably changed between fridge temperatures $10\,\rm{mK}$ and $100\,\rm{mK}$.}
\label{figure:qp_fig}
\end{figure}

We do not see any systematic temperature dependence of the quasistatic noise at low frequencies or of the Ramsey decay time $T_{\phi2}$, consistent with a picture of a paramagnetic environment with temperature-independent relaxation times (see the following sections).

\section{Fluctuation-dissipation relation between $S^\pm_\Phi(\omega)$ and magnetic susceptibility}
Here we give a simple, generic argument for how the two-sided spectrum $S^\pm_\Phi(\omega)$ is related to the dynamic susceptibility $\chi(\omega) = \chi'(\omega) + i\chi''(\omega)$ of the magnetic environment. In particular, we outline how $S_\Phi^-(\omega)$ may be related to the absorptive part of the environment's linear response, and that its $1/T$ dependence is consistent with a paramagnetic environmental susceptibility.

The most common assumption in models for $1/f$ noise is that the noise comes from a collection of dynamical fluctuators, each characterized by an exponential decay with a single relaxation time. The $1/f$ scaling then arises from an exponentially broad distribution of these relaxation times (see discussion in next section). The fluctuators could be represented by weakly interacting spin clusters of various sizes on the qubit surface, where the relaxation time is exponentially dependent on cluster size. An alternative model invokes a spin diffusion mechanism where collective diffusion modes play the role of individual fluctuators, and the broad distribution of relaxation times arises from the highly non-uniform distribution of the magnetic field around the superconducting metal of the flux qubit loop.

Whatever the microscopic source of these fluctuators may be, each fluctuator will have an effective magnetic moment operator $M_n$ that will couple flux into the qubit according to $\Delta \Phi = g_n M_n$. The two-sided flux noise spectrum in the qubit can then be written as
\begin{align}
S_\Phi(\omega) &= \int_{-\infty}^\infty d\tau e^{i\omega \tau} \langle \hat{\Phi}(\tau)\hat{\Phi}(0)\rangle\\\nonumber
&= \int_{-\infty}^\infty d\tau e^{i\omega \tau} \left\langle \sum_{n, m} g_n M_n(\tau) g_m M_m(0)\right\rangle\\\nonumber
&\approx\sum_n |g_n|^2 S_{M_{n}}(\omega),
\end{align}
where $S_{M_n}$ is the full two-sided spectral density of magnetization noise, and in the last line we have assumed negligible correlations between fluctuators (spin clusters/diffusion modes). We now use the fluctuation-dissipation theorem, which states that the equilibrium fluctuations of each moment are related to the dissipative part of its response to a non-equilibrium perturbation according to
\begin{equation}
S_{M_n}(\omega, T) = \hbar(1 + \coth{[\hbar \omega/2k_B T]})\chi''_n(\omega, T),
\end{equation}
where $\chi(\omega, T) = \chi'(\omega, T) + i\chi''(\omega, T)$ is the frequency-domain linear response function (dynamical susceptibility) to a magnetic field. Inserting this relation above yields
\begin{equation}
S_\Phi(\omega, T) = \sum_n \hbar |g_n|^2 \left[1 + \coth{\left(\frac{\hbar \omega}{2k_B T}\right)}\right]\chi''_n(\omega, T).\label{convert_to_integral}
\end{equation}
Since $\chi''(\omega,T)$ is an odd function of $\omega$, looking at the antisymmetric part shows that the experimental $1/T$ dependence of $S^-_\Phi(f)$ below the classical-quantum crossover is consistent with a dynamical susceptibility $\chi''_n$ that scales as $1/T$ for all fluctuators. This is consistent with previous measurements the $1/T$ dependence of the static susceptibility in SQUIDs \cite{Sendelbach2008}, assuming a temperature-independent distribution of relaxation times.

\section{$1/f^\alpha$ scaling near the crossover and high-frequency cutoff}
Here we discuss the $1/f^\alpha$ form of the noise and its modifications due to a possible high frequency cutoff for the relaxation times of the magnetic fluctuators. We argue that while our data clearly shows that such a cutoff must be of order $k_B T/h$ or higher, a model where the cutoff is a few times $k_B T$ fits the temperature dependence data slightly better than one with a much higher cutoff.

To obtain a $1/f^\alpha$ scaling, we assume a single relaxation time $\tau_n$ for each fluctuator. 
The dynamic susceptibility of a single such fluctuator is given by a standard Drude formula $\chi_n(\omega, T) = \frac{\chi_{n}(0, T)}{1 + i\omega\tau_n}$, meaning $\chi''_n(\omega, T) = \chi_{n}(0, T)\frac{\omega \tau_n}{1 + \omega^2 \tau_n^2}$. For paramagnetic spins, $\chi_n(0, T)\propto 1/T$. In the limit of many fluctuators with different relaxation times, we can convert (\ref{convert_to_integral}) into an integral over $\tau$ with an effective weight for each $\tau$:
\begin{equation}
S_\Phi(\omega) = \hbar\left[1 + \coth{\left(\frac{\hbar \omega}{2k_B T}\right)}\right]\int_{\tau_{\rm{min}}}^{\tau_{\rm{max}}} \rho(\tau) \frac{1}{T}\frac{\omega \tau}{1 + \omega^2\tau^2},\label{full_integration}
\end{equation}
where $\tau_{\rm{min}}$ and $\tau_{\rm{max}}$ are lower and upper cutoffs for the relaxation times, and we have included a uniform $1/T$ factor in the integrand under the assumption that all the fluctuators are paramagnetic.\footnote{We have neglected a possible dependence of the static susceptibility $\chi(0, \tau, T)$ on $\tau$ because it should be very weak. For instance, if we consider superparamagnetic clusters with different values for the total spin $S$ that fluctuate by tunneling through anisotropy barriers $U\propto S^2$, then for a given $S$, $\chi(0)\propto \frac{1}{3}S(S+1)$ while $\log\tau \propto U \propto S^2$. Thus $\chi(0, \tau, T)$ depends on $\tau$ only logarithmically and can be safely replaced by its average value.} In the classical limit,\footnote{We note that in the fully classical limit $\hbar \to 0$, $S_\Phi^-$ will vanish while $\chi''(\omega)$ (the dissipation) does not need to vanish.}
\begin{align}
S_\Phi^+(\omega\ll k_B T/\hbar) &\propto k_B \int_{\tau_{\rm{min}}}^{\tau_{\rm{max}}} d\tau \rho(\tau)\frac{\tau}{1 + \omega^2\tau^2},\label{integration}\\
S_\Phi^-(\omega\ll k_B T/\hbar) &\propto \frac{\hbar}{T}\int_{\tau_{\rm{min}}}^{\tau_{\rm{max}}} d\tau \rho(\tau)\frac{\omega \tau}{1 + \omega^2\tau^2}.
\end{align}
As before, from this we can see that our experimental data below the classical-quantum crossover is consistent with an environment of magnetic fluctuators with a paramagnetic static susceptibility $\chi(0)$, under the assumption that $\rho(\tau)$ is independent of temperature.

\begin{figure}[t!]
\begin{centering}
\includegraphics[width=.9\linewidth]{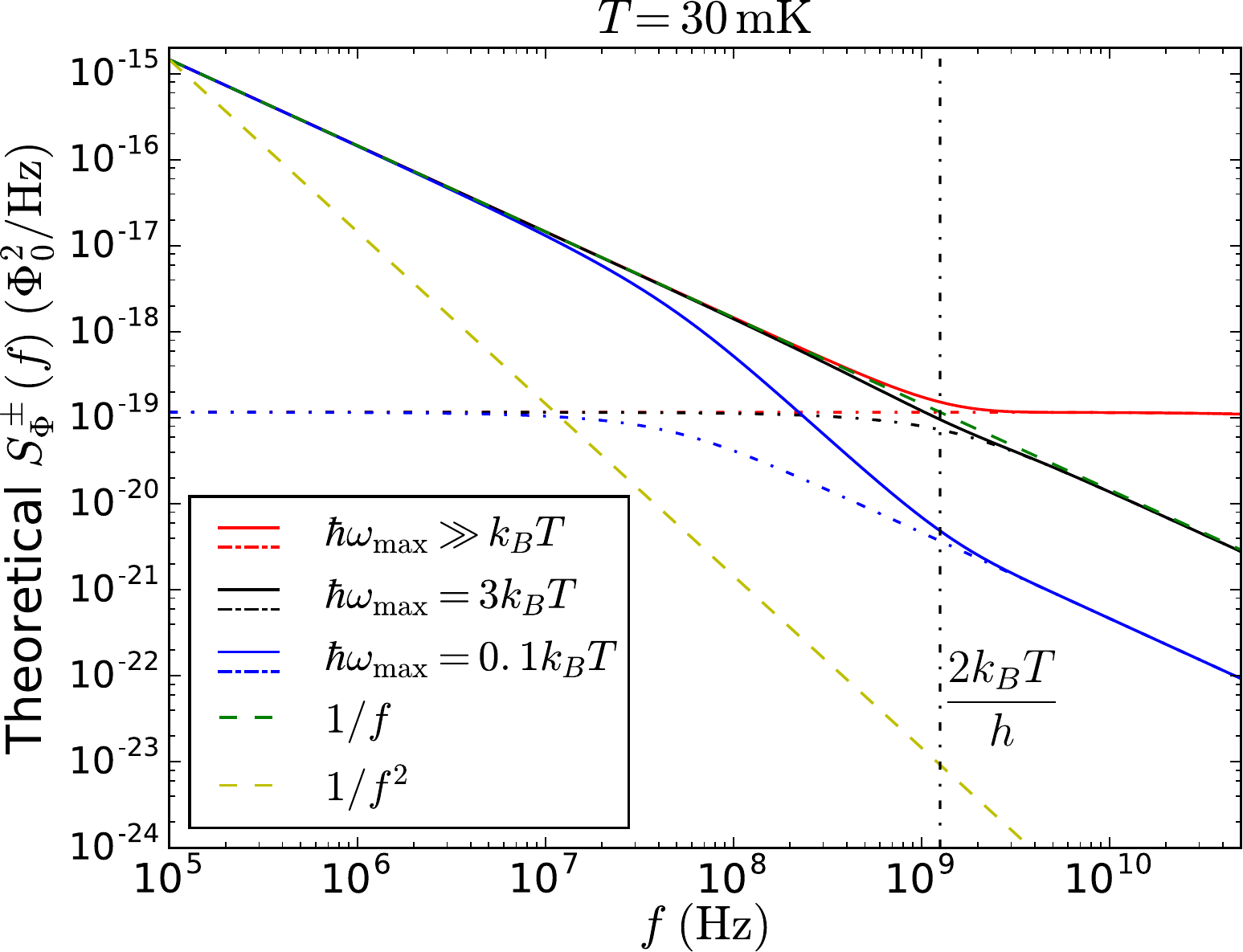}
\par\end{centering}
\caption{\footnotesize $1/f$ noise according to the finite high frequency cutoff model (\ref{high_freq_cutoff_model}) for different values of $\omega_{\rm{max}}$, showing the three qualitative `flavors' the noise scaling near the classical-quantum crossover. Dash-dotted lines are $S^-_\Phi$ while solid lines are $S^+_\Phi$. The `smoothest' transition through the cutoff where $S^+_\Phi(f)$ remains close to $1/f$ for all frequencies is achieved for $\omega_{\rm{max}}\approx 3 k_B T/\hbar$.}
\label{figure:cutoffs}
\end{figure}

If we postulate $\rho(\tau)\propto \frac{1}{\tau}$ between $\tau_1$ and $\tau_2$, then performing the integration leads to $S_\Phi^+(\omega)\propto 1/\omega$ assuming $\frac{1}{\tau_2} \ll \omega \ll \frac{1}{\tau_1}$ ($\omega_{\rm{min}} \ll \omega \ll \omega_{\rm{max}}$), which is the usual picture of $1/f$ noise in the classical limit. However, more precisely performing the full integration (\ref{full_integration}) without assuming anything about $\omega$ relative to $T$ or $\tau_{\rm{min/max}}$ yields for the full spectrum
\begin{equation}
S_\Phi(\omega) \propto \frac{\hbar}{T}\left[1 + \coth{\left(\frac{\hbar \omega}{2k_B T}\right)}\right] \left.\tan^{-1}(\omega \tau)\right|^{\tau_{\rm{max}}}_{\tau_{\rm{min}}}\label{high_freq_cutoff_model}
\end{equation}
As long as $\tau_{\rm{min/max}}$ don't depend exponentially on $T$, the temperature dependence of $S_\Phi^-$ will again be given by that of the static susceptibility. Let us assume that $\omega_{\rm{min}} \ll k_B T/\hbar$ (justified by the presence of $1/f$ flux noise well below $1\,\rm{GHz}$) and look at the shape of the classical-quantum crossover for different $\omega_{\rm{max}}$. Fig. \ref{figure:cutoffs} shows that there are three qualitative types of scaling behavior of $S^\pm_\Phi(f)$ around the crossover. If $\omega_{\rm{max}}\ll k_B T/\hbar$, then $S^+_\Phi(f)$ would approach the crossover from below as $1/f^2$, inconsistent with the data. On the other hand, if $\omega_{\rm{max}} \gg k_B T/\hbar$, then $S^+_\Phi(f)$ will turn into white noise just above the crossover. There is also an intermediate regime $\omega_{\rm{max}} \approx 3 k_B T/\hbar$ where $S^+_\Phi(f)$ is very close to $1/f$ for all frequencies except for a slight deviation at the crossover point.

\begin{figure}[h!]
\begin{centering}
\includegraphics[width=1.0\linewidth]{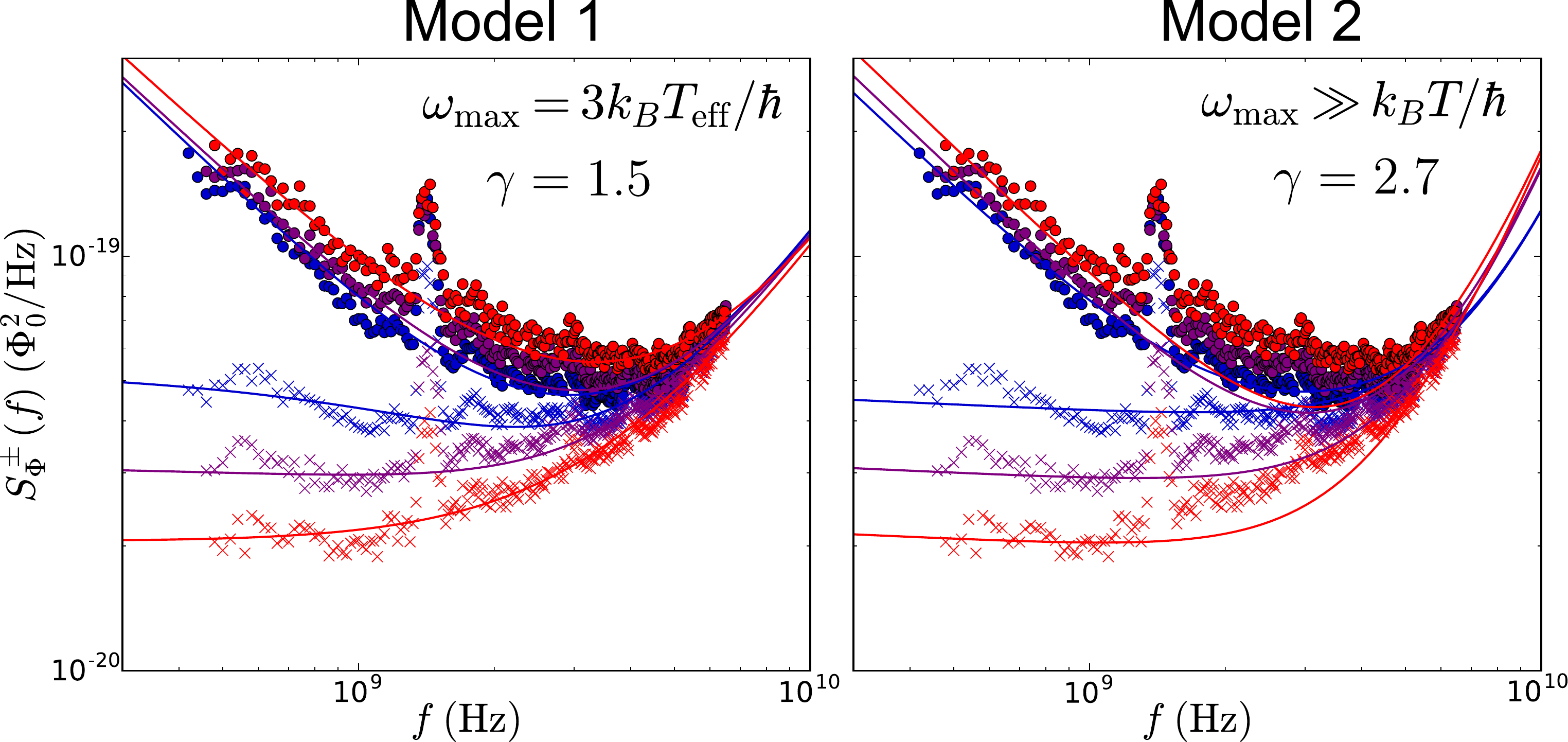}
\par\end{centering}
\caption{\footnotesize Comparison of two models with the constraint that the ohmic exponent $\gamma$ be the same at all temperatures. The model with a finite, temperature-dependent $\omega_c \approx 3 k_B T_{\rm{eff}}$ fits the temperature dependence data somewhat better than the phenomenological model with $\alpha = 1.05$ used in the main text at base temperature. Here, $T_{\rm{eff}}$ is the temperature deduced from the data below the crossover. However, there is no significant difference between the fits just looking at $30\,\rm{mK}$ alone.}
\label{figure:compare_models}
\end{figure}

In Fig. \ref{figure:compare_models}, we compare the phenomenological thermodynamic power law model used in the main text,
\begin{align}
S^{\rm{phen.}}_\Phi(\omega) &= A\frac{\omega}{|\omega|^{\alpha}} [1 + \coth{(\hbar \omega/[2 k_B T_A])}]\\ &+ B \omega |\omega|^{\gamma-1} [1 + \coth{(\hbar \omega/[2 k_B T_B])}],
\end{align}
to the finite high-frequency cutoff model (\ref{high_freq_cutoff_model}). The former implicitly assumes that $\omega_{\rm{max}} \gg k_B T/\hbar$, so that any deviation of $S_\Phi^+$ from a perfect $1/f$ scaling near the classical-quantum crossover would be due to temperature alone. Although this seems to fit our data at base temperature, we find that incorporating a finite $\omega_{\rm{max}}$ fits the higher temperature data slightly better under the conditions of the fit (constant $\alpha$ and $\gamma$ for all temperatures). Namely, we use the ``intermediate'' type of crossover, $\omega_{\rm{max}} = 3k_B T_{\rm{eff}}$, where $T_{\rm{eff}}$ is the limiting effective temperature deduced from the stray population data below the crossover. A high frequency cutoff that scales linearly with temperature might have a natural physical meaning. For instance, if the inverse relaxation times $\tau_n^{-1}$ are determined by spin-phonon interactions, both $\tau_n^{-1}$ and $\omega_{\text{max}}$ will be proportional to the number of phonons with energy close to the typical Zeeman splittings of the fluctuators, which scales linearly with $T$ (since the Zeeman splittings should be $\ll k_B T$ even for clusters).

\section{Implications of high frequency cutoff for spin diffusion}
We mention one more mechanism that would give a frequency-dependent $\alpha$ due to a finite $\omega_{\rm{max}}$, but with a different functional form. Namely, we consider spin diffusion, which was proposed by Faoro and Ioffe \cite{Faoro2008} to explain $1/f$ noise in SQUIDs and further explored in the context of D-Wave flux qubits by Lanting \emph{et al.} \cite{Lanting2014}. We conclude that given the $1/f$ scaling of $S^+_\Phi(f)$ near $1\,\rm{GHz}$, spin diffusion is unlikely to be the source of the $1/f$ noise near the classical-quantum crossover unless i.) the spin diffusion constant is several orders of magnitude higher than estimates in the literature \cite{Faoro2008, Lanting2014} or ii.) the spin density is substantially inhomogeneous, leading to a shallower power law at high frequencies, in which case a separate physical mechanism needs to be invoked for the $1/f$ power law at lower frequencies.

Within the spin diffusion model, the total spin is conserved and spin excitations will diffuse around the surface of the superconducting qubit metal, generating flux noise by coupling to a non-uniform distribution of the magnetic field. In Ref. \cite{Faoro2008}, the mechanism of diffusion was proposed to be an RKKY interaction mediated through the superconductor. The diffusion equation for the coarse-grained magnetization is
\begin{equation}
\partial M_\alpha(\bm{r}, t)/\partial t = \mathcal{D} \nabla^2 M_\alpha(\bm{r}, t),\label{diffusion}
\end{equation}
where $\mathcal{D}$ is the diffusion coefficient and $\alpha=x,y,z$. Eq. (\ref{diffusion}) can be solved using the Laplace transform $M_\alpha(\bm{r}) \propto e^{-\Gamma_n t}\varphi_n(\bm{r})$, which leads to the eigenvalue problem
\begin{equation}
\nabla^2\varphi_n(\bm{r})= -\Gamma_n\varphi_n(\bm{r}),
\end{equation}
with periodic boundary conditions on the surface of the qubit metal. Solving for the eigenmodes $\varphi_n(\bm{r})$ allows one to express the dynamic magnetic susceptibility from the Green's function for Eq. (\ref{diffusion}):
\begin{equation}
\chi_{\alpha \beta}''(\bm{r}, \bm{r}', \omega) = \delta_{\alpha\beta} \chi(0,T)\sum_n\varphi_n(\bm{r})\frac{\omega \Gamma_n}{\omega^2 + \Gamma_n^2}\varphi_n(\bm{r}').
\end{equation}
The flux noise then becomes a sum of Lorentzians corresponding to each diffusion mode
\begin{equation}
S_\Phi(\omega) = \hbar \omega \chi(0, T)\left[1 + \coth{\left( \frac{\hbar \omega}{2k_B T}\right)} \right]\sum_n b_n^2\frac{\Gamma_n}{\omega^2 + \Gamma_n^2},
\end{equation}
where $b_n^2 = \sum_\alpha b^2_{\alpha n}$, $b_{\alpha n}=\int d\bm{r}b_\alpha(\bm{r})\varphi_n(\bm{r})$, are coupling factors describing how the local moment couples to the qubit's magnetic field and by reciprocity the qubit loop itself. Therefore the spin diffusion model naturally leads to a broad set of relaxation times $\tau_n = 1/\Gamma_n$ with a temperature-independent distribution function $\rho(\tau_n)=b_n^2$ given by the form factors. The temperature dependence of $S_\Phi^-$ will then arise solely from that of the static magnetic susceptibility $\chi(0, T)$.

In the limit of a flat wire (thin film), one would expect a $1/f^{\alpha = 1}$ power law for $\omega_{\rm{min}} \ll \omega \ll \omega_{\rm{max}}, k_B T/
\hbar$. But allowing for a finite aspect ratio or for inhomogeneity of the spin density on the surface, one can have an exponent $\alpha\ne 1$. In particular, if the spins are concentrated near edges we can have $\alpha < 1$ \cite{PetukhovUnpublished}. An analytic approximation for the noise summation is \cite{PetukhovUnpublished}
\begin{equation}
S_\Phi(\omega) = A\left[1 + \coth{\left( \frac{\hbar \omega}{2k_B T}\right)} \right] \frac{\hbar \omega}{k_B T}\int_0^\infty dx\,\frac{x^{3 - 2\alpha}e^{-\frac{x}{\sqrt{\omega_{\rm{max}}}}}}{\omega^2 + x^4}.
\end{equation}

One would expect the high frequency cutoff $\omega_{\rm{max}}$ for $\Gamma_n$ to be given by $\omega_{\rm{max}}\approx \mathcal{D}/\ell^2$, where $\ell$ is the smallest dimension associated with the qubit geometry, which for our device should be the metal thickness of $\sim 100\,\rm{nm}$. The highest estimates in the literature for $\mathcal{D}$ are $10^8-10^9\,\rm{nm}^2/s$ \cite{Faoro2008}, which would imply a physically expected cutoff of $\omega_{\rm{max}}/(2\pi)\sim 100\,\rm{kHz}$.\footnote{Conversely, a low-frequency cutoff of $\sim 10\,\rm{Hz}$ would be expected, which would mean the noise below $1\,\rm{Hz}$ would need to have a different physical mechanism, unless the diffusion constant is much lower than expected. The ``bump'' in flux noise observed at intermediate frequencies in the main paper could potentially be due to spin diffusion, assuming it is not suppressed by spin relaxation.} Therefore, for spin diffusion to be relevant to the classical-quantum crossover, we would need either a much larger $\mathcal{D}$ and/or $\alpha < 1$. If $\alpha < 1$, this would mean the quasistatic flux noise (which shows $\alpha = 1$ below $10\,\rm{Hz}$) is not spin diffusion noise, but it may be possible to have different physical mechanisms in different frequency ranges with the noise power still scaling similarly between samples at high and low frequencies.

%